# Analysis of the Fractional Relativistic Polytropic Gas Sphere

Mohamed S. Aboueisha[1], Mohamed I. Nouh[1,*], Eamad A-B. Abdel-Salam[2]. Tarek M. Kamel[1], Mohamed M. Beheary[3] and Kamel A. K. Gadallah[3]

[1]Astronomy Department, National Research Institute of Astronomy and Geophysics, 11421 Helwan, Cairo, Egypt

[2]Department of Mathematics, Faculty of Science, New Valley University, El-Kharja 72511, Egypt

[3]Department of Astronomy and Meteorology, Faculty of Science Al-Azhar University, Nasr City, 11889, Cairo, Egypt

mohamed.nouh@nriag.sci.eg

**Abstract:** Many stellar configurations, including white dwarfs, neutron stars, black holes, supermassive stars, and star clusters, rely on relativistic effects. The Tolman-Oppenheimer-Volkoff (TOV) equation of the polytropic gas sphere is ultimately a hydrostatic equilibrium equation developed from the general relativity framework. In the modified Riemann Liouville (mRL) frame, we formulate the fractional TOV (FTOV) equations and introduce an analytical solution. Using power series expansions in solving fractional TOV equations yields a limited physical range to the convergent power series solution. Therefore, combining the two techniques of Euler-Abel transformation and Padé approximation has been applied to improve the convergence of the obtained series solutions. For all possible values of the relativistic parameters ($\sigma$), we calculated twenty fractional gas models for the polytropic indexes n=0, 0.5, 1, 1.5, 2. Investigating the impacts of fractional and relativistic parameters on the models revealed fascinating phenomena; the two effects for n=0.5 are that the sphere's volume and mass decrease with increasing $\sigma$ and the fractional parameter ($\alpha$). For n=1, the volume decreases when $\sigma=0.1$ and then increases when $\sigma=0.2$ and 0.3. The volume of the sphere reduces as both $\sigma$ and $\alpha$ increase for n=1.5 and n=2. We calculated the maximum mass and the corresponding minimum radius of the white dwarfs modeled with polytropic index n=3 and several fractional and relativistic parameter values. We obtained a mass limit for the white dwarfs somewhat near the Chandrasekhar limit for the integer models with small relativistic parameters ($\alpha = 1$, $\sigma = 0.001$). The situation is altered by lowering the fractional parameter; the mass limit increases to $M_{limit}$=1.63348 M at $\alpha = 0.95$ and $\sigma = 0.001$.

## 1. Introduction

Relativistic effects play an essential role in compact stars, supermassive stars, and star clusters [1-5]. The Tolman-Oppenheimer-Volko (TOV) equations describe the gravitational potential of the self-gravitating gas sphere as a function of radius in curved spacetime. These equations and the equation of state form a closed set of nonlinear differential equations, usually solved by numerical integration techniques [6-7]. [8] analyzed a mass function that they claim gives a solution to the TOV equation for an isotropic and spherically symmetric system using the homotopy perturbation approach.

Relativistic studies of the polytropic equation of state, achieved by [9,6,10], present an approximate analytical solution for the polytropic indices $n$ illustrating the results for the relativistic parameters $\sigma$.

When solving the differential equation (like the polytropic LE equation) using a power series, the main aim is to ensure that the series converges to the outer surface of a gas sphere. [11] used Euler transformation to accelerate the series to LE equation of the polytropic and isothermal gas spheres in this context. [12] used two accelerating techniques to improve the radius of convergence for the range of the polytropic index $0 < n < 5$.

Physical science, astrophysics, anomalous diffusion, signal processing, and quantum mechanics all benefit from modeling with fractional differential equations. [13] examined the second law of thermodynamics for the Friedmann Universe bounded by a boundary in the framework of fractional action cosmology. [14] used a power law weight function to create a dark energy model in fractional action cosmology. [15] examined the fractional white dwarf model, while [16] used the Taylor series to solve the fractional isothermal gas sphere. Several studies by [17-19] developed solutions to the polytropic and isothermal gas spheres using power series expansion.

In the modified Riemann Liouville (mRL) frame, we introduce a new analytic solution to the fractional relativistic polytropic gas sphere using the series expansion method. The construct of the solution rests mainly on accelerating the power series solution of the factional TOV equations using a combination of two different transformations. Moreover, our calculations are presented graphically for fractional gas models of the polytropic indexes $n = 0, 0.5, 1, 1.5, 2, 3$, covering all possible values of the relativistic parameter. We analyse the behaviour of the fractional mass function and the fractional radius of the polytrope regarding the integer ones. Moreover, the detailed calculation for polytrope with index n=3 will be performed to calculate the maximum mass limit and its corresponding minimum radius of mRL fractional white dwarfs.

The structure of the paper is as follows: Section 2 is devoted to the principles of the mRL derivatives.; in Section 3, we formulated the fractional TOV equations; Section 4 deals with the solution of the FTOV equation; the results of the analysis are summarized in section 5, and section 5 present the conclusion.

## 2. The mRL Fractional Derivative

The mRL derivative is introduced by [20-24]. Suppose that $f : R \rightarrow R$ where $x \rightarrow f(x)$ is a continuous function and $h$ denotes a constant discretization span. The mRL derivative is given in the form

$$f^{(\alpha)}(x) = \lim_{h \to \infty} \frac{\Delta^{\alpha}\left[f(x) - f(0)\right]}{h^{\alpha}}, \qquad 0 < \alpha < 1 \tag{1}$$

where

$$\Delta^{\alpha} f(x) = \sum_{k=0}^{\infty} (-1)^{k} \frac{\Gamma(\alpha+1)}{\Gamma(k+1)\Gamma(\alpha-k+1)} f\left[x + (\alpha-k)h\right], \tag{2}$$

where $\Gamma(x)$ represents the Gamma function.

This is similar to the standard of derivatives, where the $\alpha$ order derivative of a constant is zero.

$$D^{\alpha} f(x) = \begin{cases} \dfrac{1}{\Gamma(-\alpha)} \int_{0}^{x} (x-\xi)^{-\alpha-1} \left[f(\xi) - f(0)\right] d\xi & \alpha < 0 \\[2mm] \dfrac{1}{\Gamma(1-\alpha)} \dfrac{d}{dx} \int_{0}^{x} (x-\xi)^{-\alpha} \left[f(\xi) - f(0)\right] d\xi & 0 < \alpha < 1 \\[2mm] \dfrac{1}{\Gamma(n-\alpha)} \dfrac{d^{n}}{dx^{n}} \int_{0}^{x} (x-\xi)^{n-\alpha-1} \left[f(\xi) - f(0)\right] d\xi & n < \alpha < n+1, n \geq 1 \end{cases} \tag{3}$$

Some useful Jumarie-modified formulae could be summarized as

$$D_{x}^{\alpha} x^{\gamma} = \frac{\Gamma(\gamma+1)}{\Gamma(\gamma-\alpha+1)} x^{\gamma-\alpha}, \qquad \gamma > 0 \tag{4}$$

$$D_{x}^{\alpha}\left(cf(x)\right) = c D_{x}^{\alpha} f(x), \tag{5}$$

$$D_{x}^{\alpha}\left[f(x) g(x)\right] = g(x) D_{x}^{\alpha} f(x) + f(x) D_{x}^{\alpha} g(x), \tag{6}$$

$$D_x^\alpha f[g(x)] = f_g'[g[x]] D_x^\alpha g(x), \qquad (7)$$

$$D_x^\alpha f[g(x)] = D_g^\alpha f[g[x]](g_x')^\alpha, \qquad (8)$$

Since $c$ is a constant.

Equations (6) to (8) are direct results from

$$D_x^\alpha f(x) \cong \Gamma(\alpha+1) D_x f(x), \qquad (9)$$

Following [25], equation (7) could be modified to

$$D_x^\alpha f[g(x)] = \omega_x f_g'[g[x]] D_x^\alpha g(x) \qquad (10)$$

Consequently, Equations (6) and (8) will be modified to the following forms

$$D_x^\alpha [f(x) g(x)] = \omega_x \{g(x) D_x^\alpha f(x) + f(x) D_x^\alpha g(x)\}, \qquad (11)$$

$$D_x^\alpha f[g(x)] = \omega_x D_g^\alpha f[g[x]](g_x')^\alpha, \qquad (12)$$

where $\omega_x$ is called the fractal index determined in terms of gamma functions as follows [26]

$$\omega_x = \frac{\Gamma(m\alpha+1)}{m\Gamma(m\alpha+1)\Gamma(m\alpha-\alpha+1)} . \qquad (13)$$

The fractional parameter $\alpha$ takes values less or equal to 1 in our calculations. The reason for this choice is that, as we know, the physical meaning of the first-order derivative is the velocity or speed, the second derivative is the acceleration, the third order is the jerk, and the fourth order is the snap or jounce, so we can define the first fractional derivative as the spatial kind of velocity and become the well-known velocity when $\alpha=1$. Also, $D_t^\alpha(D_t^\alpha r) = D_t^{\alpha\alpha} r$, $0 < \alpha \leq 1$ is the spatial kind of acceleration and return to the well-known acceleration when $\alpha=1$, and so. But if one chooses, $\alpha > 1$ say $1 < \alpha \leq 2$, we will assume for the sake of argument that it represents the acceleration, but in fact, it does not represent the acceleration because we cannot determine the speed that led to that acceleration, and therefore, the values that are larger than one have no physical meaning.

### 3.    Formulation of FTOV Equations

In the case of spherically symmetric (r, $\vartheta$, $\varphi$) and static spacetime (t), the line element signature is given by [27-28]

$$ds^2 = e^{2v(r)}(cd\,t)^2 - e^{2\lambda(r)}d\,r^2 - r^2 d\,\vartheta^2 - r^2 \sin^2(\vartheta)d\,\varphi^2 \tag{14}$$

We can write Equation (14) in the fractional form as

$$d^\alpha s^2 = e^{2v}(cd^\alpha t)^2 - e^{2\lambda}d^\alpha r^2 - r^{2\alpha}d^\alpha \vartheta^2 - r^{2\alpha}\sin^2(\vartheta^\alpha)d^\alpha \varphi^2, \tag{15}$$

Where $v$ and $\lambda$ are functions of $r^\alpha$ only and governed by Einstein's field equation. The relation between the line element and the metric tensor $^\alpha g_{\mu\nu}$ is

$$d^\alpha s^2 = {}^\alpha g_{\mu\nu}d^\alpha x^\mu d^\alpha x^\nu. \tag{16}$$

By comparing Equations (15) and (16), the most general metric tensor in the fractional form is

$$^\alpha g_{\mu\nu} = \begin{pmatrix} e^{2v} & 0 & 0 & 0 \\ 0 & -e^{2\lambda} & 0 & 0 \\ 0 & 0 & -r^{2\alpha} & 0 \\ 0 & 0 & 0 & -r^{2\alpha}\sin^2(\vartheta^\alpha) \end{pmatrix}, \tag{17}$$

from this, the nonzero elements of Equation (17) are

$$^\alpha g_{tt} = {}^\alpha g_{11} = e^{2v}, \qquad ^\alpha g_{rr} = {}^\alpha g_{22} = -e^{2\lambda},$$
$$^\alpha g_{\vartheta\vartheta} = {}^\alpha g_{33} = -r^{2\alpha}, \qquad ^\alpha g_{\varphi\varphi} = {}^\alpha g_{44} = -r^{2\alpha}\sin^2\vartheta^\alpha. \tag{18}$$

The fractional form of the curvature tensor is defined in terms of Christoffel symbols as

$$^\alpha R^\lambda_{\beta\mu\nu} = {}^\alpha \Gamma^\lambda_{\nu\beta,\mu} - {}^\alpha \Gamma^\lambda_{\mu\beta,\nu} + {}^\alpha \Gamma^a_{\beta\nu}{}^\alpha \Gamma^\lambda_{\mu a} - {}^\alpha \Gamma^a_{\mu\beta}{}^\alpha \Gamma^\lambda_{\nu a}, \tag{19}$$

where the comma subscript notation represents the coordinate's fractional derivative, i.e.

$${}^{\alpha}S,_{\mu} = \frac{\partial^{\alpha}S}{\partial x^{\mu\alpha}} \text{ or simply } {}^{\alpha}S,_{\mu} = D_{\mu}^{\alpha}S.$$

The curvature at a point in space is fully described by

$${}^{\alpha}\Gamma_{\mu\nu}^{\lambda} = \frac{1}{2} {}^{\alpha}g^{\lambda a} ({}^{\alpha}g_{a\nu,\mu} + {}^{\alpha}g_{a\mu,\nu} - {}^{\alpha}g_{\mu\nu,a}). \tag{20}$$

Note that

$${}^{\alpha}g^{tt} = {}^{\alpha}g^{11} = \frac{1}{e^{2\nu}}, \qquad {}^{\alpha}g^{rr} = {}^{\alpha}g^{22} = -\frac{1}{e^{2\lambda}},$$

$${}^{\alpha}g^{\vartheta\vartheta} = {}^{\alpha}g^{33} = -\frac{1}{r^{2\alpha}}, \qquad {}^{\alpha}g^{\varphi\varphi} = {}^{\alpha}g^{44} = -\frac{1}{r^{2\alpha}\sin^{2}\vartheta^{\alpha}}. \tag{21}$$

So, we can calculate some of the fractional Christoffel symbols, ${}^{\alpha}\Gamma_{tt}^{t}$, ${}^{\alpha}\Gamma_{tt}^{r}$, ${}^{\alpha}\Gamma_{rt}^{t}$, ${}^{\alpha}\Gamma_{rr}^{r}$, ${}^{\alpha}\Gamma_{\theta\theta}^{r}$, ${}^{\alpha}\Gamma_{\phi\phi}^{r}$, ${}^{\alpha}\Gamma_{r\theta}^{\theta}$, ${}^{\alpha}\Gamma_{r\phi}^{\phi}$, ${}^{\alpha}\Gamma_{\theta\phi}^{\phi}$, and ${}^{\alpha}\Gamma_{\phi\phi}^{\theta}$.

The Ricci tensor is obtained from the Riemann tensor by contracting over two of the indices.

$${}^{\alpha}R_{\mu\nu} \equiv {}^{\alpha}R_{\mu\nu k}^{k} = {}^{\alpha}\Gamma_{\mu\nu,k}^{k} - {}^{\alpha}\Gamma_{\nu k,\mu}^{k} + {}^{\alpha}\Gamma_{\mu\nu}^{k} {}^{\alpha}\Gamma_{km}^{m} - {}^{\alpha}\Gamma_{\mu m}^{k} {}^{\alpha}\Gamma_{\nu k}^{m}. \tag{22}$$

The Ricci scalar is then given by

$${}^{\alpha}R = {}^{\alpha}g^{\mu\nu} {}^{\alpha}R_{\mu\nu}. \tag{23}$$

In fractional spacetime, the fractional Einstein tensor ${}^{\alpha}G_{\mu\nu}$ can be written down by using the fractional metric tensor ${}^{\alpha}g_{\mu\nu}$, the fractional Ricci tensor ${}^{\alpha}R_{\mu\nu}$, and the fractional Ricci scalar ${}^{\alpha}R$ as

$${}^{\alpha}G_{\mu\nu} \equiv {}^{\alpha}R_{\mu\nu} - \frac{1}{2} {}^{\alpha}R {}^{\alpha}g_{\mu\nu} = \frac{8\pi G}{c^{4}} {}^{\alpha}T_{\mu\nu}. \tag{24}$$

The fractional Christoffel symbols for spherically symmetric sources with coordinates $(ct, r, \theta, \phi)$ are

$$
\begin{aligned}
&{}^{\alpha}\Gamma^{r}_{tt} = {}^{\alpha}\Gamma^{2}_{11} = e^{2(\nu-\lambda)}D^{\alpha}_{r}\nu, &&{}^{\alpha}\Gamma^{t}_{tt} = {}^{\alpha}\Gamma^{1}_{11} = 0, \\
&{}^{\alpha}\Gamma^{t}_{rt} = {}^{\alpha}\Gamma^{1}_{21} = D^{\alpha}_{r}\nu, &&{}^{\alpha}\Gamma^{r}_{rr} = {}^{\alpha}\Gamma^{2}_{22} = -D^{\alpha}_{r}\lambda, \\
&{}^{\alpha}\Gamma^{r}_{\vartheta\vartheta} = {}^{\alpha}\Gamma^{2}_{33} = -\frac{Z r^{\alpha} e^{-2\lambda}}{2}, &&{}^{\alpha}\Gamma^{r}_{\phi\phi} = {}^{\alpha}\Gamma^{2}_{44} = -\frac{Z r^{\alpha}\sin^{2}\vartheta^{\alpha} e^{-2\lambda}}{2}, \\
&{}^{\alpha}\Gamma^{\theta}_{r\theta} = {}^{\alpha}\Gamma^{3}_{23} = \frac{Z}{2r^{\alpha}}, &&{}^{\alpha}\Gamma^{\phi}_{r\phi} = {}^{\alpha}\Gamma^{4}_{24} = \frac{Z}{2r^{\alpha}}, \\
&{}^{\alpha}\Gamma^{\phi}_{\theta\phi} = {}^{\alpha}\Gamma^{4}_{34} = \frac{Z}{2r^{\alpha}}\cot\theta^{\alpha}, &&{}^{\alpha}\Gamma^{\theta}_{\phi\phi} = {}^{\alpha}\Gamma^{3}_{44} = -\frac{Z}{2}\sin\theta^{\alpha}\cos\theta^{\alpha},
\end{aligned}
\tag{25}
$$

where $Z = \dfrac{\Gamma(2\alpha+1)}{\Gamma(\alpha+1)}$.

Note that ${}^{\alpha}\Gamma^{\lambda}_{\mu\nu} = {}^{\alpha}\Gamma^{\lambda}_{\nu\mu}$, and all symbols not specified are zero or related by symmetry relations. Using the previous declarations, one can calculate the nonzero independent components of the fractional Ricci tensor ${}^{\alpha}R_{\mu\nu}$ for a spherically symmetric source by using Equations (22) and (25)

$$
\begin{aligned}
&{}^{\alpha}R_{tt} = {}^{\alpha}R_{11} = \left(D^{\alpha\alpha}_{r}\nu - D^{\alpha}_{r}\nu D^{\alpha}_{r}\lambda + (D^{\alpha}_{r}\nu)^{2} + \frac{Z}{r^{\alpha}}D^{\alpha}_{r}\nu\right)e^{2(\nu-\lambda)}, \\
&{}^{\alpha}R_{rr} = {}^{\alpha}R_{22} = D^{\alpha\alpha}_{r}\nu - D^{\alpha}_{r}\nu D^{\alpha}_{r}\lambda + (D^{\alpha}_{r}\nu)^{2} + \frac{Z}{r^{\alpha}}D^{\alpha}_{r}\lambda, \\
&{}^{\alpha}R_{\theta\theta} = {}^{\alpha}R_{33} = \left(1 + \frac{Z}{2}r^{\alpha}D^{\alpha}_{r}\nu - \frac{Z}{2}r^{\alpha}D^{\alpha}_{r}\lambda\right)e^{-2\lambda} - 1, \\
&{}^{\alpha}R_{\phi\phi} = {}^{\alpha}R_{44} = {}^{\alpha}R_{22}\sin^{2}\theta^{\alpha}.
\end{aligned}
\tag{26}
$$

The fractional scalar curvature represented by Equation (23) could be written as a contraction of the fractional Ricci tensor with the fractional metric

$$
{}^{\alpha}R = -\frac{2e^{-2\lambda}}{r^{2\alpha}}\left(\begin{array}{l} r^{2\alpha}D^{\alpha\alpha}_{r}\nu - r^{2\alpha}D^{\alpha}_{r}\nu D^{\alpha}_{r}\lambda + r^{2\alpha}(D^{\alpha}_{r}\nu)^{2} \\ + Z\{r^{\alpha}D^{\alpha}_{r}\nu - r^{\alpha}D^{\alpha}_{r}\lambda\} - e^{2\lambda} + 1 \end{array}\right).
\tag{27}
$$

The fractional Einstein's curvature tensor, Equation (24), can now be constructed from the fractional Ricci tensor and the fractional scalar curvature as

$$
{}^{\alpha}G_{(t)(t)} = \frac{e^{-2\lambda}}{r^{2\alpha}}(Zr^{\alpha}D^{\alpha}_{r}\lambda - 1) + \frac{1}{r^{2\alpha}} = \frac{8\pi G}{c^{4}}\rho c^{2},
\tag{28}
$$

$$
{}^{\alpha}G_{(r)(r)} = \frac{e^{-2\lambda}}{r^{2\alpha}}(1 + Zr^{\alpha}D^{\alpha}_{r}\nu) - \frac{1}{r^{2\alpha}} = \frac{8\pi G}{c^{4}}p,
\tag{29}
$$

$$^{\alpha}G_{(\theta)(\theta)} = e^{-2\lambda}\left(D_r^{\alpha\alpha}v + (D_r^{\alpha}v)^2 - D_r^{\alpha}\lambda D_r^{\alpha}v + \frac{Z}{2r^{\alpha}}(D_r^{\alpha}\lambda - D_r^{\alpha}v)\right),$$
$$= \frac{8\pi G}{c^4}p \tag{30}$$

$$^{\alpha}G_{(\phi)(\phi)} = {}^{\alpha}G_{(\theta)(\theta)}. \tag{31}$$

The fractional energy-momentum tensor is symmetric and contains all relevant information about mass density, energy density, momentum density, etc. The fractional energy-momentum tensor is given in terms of pressure $p$ and the mass density $\rho$ of a given stellar object. We will assume the matter to be described by a perfect fluid at rest in the case of a static, spherically symmetric fractional metric so that

$$^{\alpha}T^{\mu\nu} = \rho\,{}^{\alpha}u^{\mu}\,{}^{\alpha}u^{\nu} + \frac{p}{c^2}\left({}^{\alpha}u^{\mu}\,{}^{\alpha}u^{\nu} - {}^{\alpha}g^{\mu\nu}\right). \tag{32}$$

Where $\rho$ is the proper mass density and $p$ isotropic pressure in the rest frame of the fluid, both may be taken as functions only of the radial coordinate r for a static matter distribution and $^{\alpha}u^{\mu}$ the fractional velocity vector.

By the law of local energy-momentum conservation $D_{\nu}^{\alpha}\,{}^{\alpha}T^{\mu\nu} = 0$, the r component of this conservation law is

$$(\rho c^2 + p)D_r^{\alpha}v = -D_r^{\alpha}p. \tag{33}$$

Which is the fractional hydrostatic equilibrium equation that describes the balance between gravitational force and pressure gradient. We'll start with the (t)(t) component (Equation (26))

$$^{\alpha}G_{(t)(t)} = \frac{1}{r^{2\alpha}} - \frac{e^{-2\lambda}}{r^{2\alpha}} - \frac{1}{r^{\alpha}}D_r^{\alpha}e^{-2\lambda} = \frac{8\pi G}{c^2}\rho, \tag{34}$$

where $G$ is the gravitational acceleration and $c$ is the light velocity.
The last equation can be transferred into the form

$$D_r^{\alpha}\left(r^{\alpha}(1-e^{-2\lambda})\right) = D_r^{\alpha}\left(\frac{ZG}{c^2}m(r^{\alpha})\right),$$

where $Z = \Gamma(2\alpha+1)/\Gamma(\alpha+1)$ and the mass contained in the radius $r^{\alpha}$ is given by

$$m(r^\alpha) = \int_0^{r^\alpha} 4\pi x^\alpha \rho(x^\alpha) dx^\alpha ,  \tag{35}$$

we arrive at the relation

$$e^{2\lambda} = \left[1 - \frac{ZG\,m(r^\alpha)}{c^2 R}\right]^{-1} .  \tag{36}$$

The (r)(r) component (Equation (26)) of the field equations reads

$$^\alpha G_{(r)(r)} = -\frac{1}{r^{2\alpha}} + \frac{Ze^{-2\lambda}}{r^\alpha} D_r^\alpha v = \frac{8\pi G}{c^4} p ,$$

we obtain the relation

$$D_r^\alpha v = \frac{\dfrac{ZG}{c^2} m(r^\alpha) + \dfrac{8\pi G}{c^4} p r^{3\alpha}}{Z r^\alpha \left[r^\alpha - \dfrac{ZG}{c^2} m(r^\alpha)\right]} .  \tag{37}$$

This enables us to put the fractional equation of hydrostatic equilibrium into the fractional Tolman-Oppenheimer-Volko (FTOV) form as

$$D_r^\alpha p = -(\rho c^2 + p) \frac{\dfrac{ZG}{c^2} m(r^\alpha) + \dfrac{8\pi G}{c^4} p r^{3\alpha}}{Z r^\alpha \left[r^\alpha - \dfrac{2G}{c^2} m(r^\alpha)\right]} ,$$

and in another form is

$$D_r^\alpha p = -\frac{GM\rho}{r^{3\alpha}} \left[1 + \frac{p}{c^2 \rho}\right] \left[1 + \frac{4\pi r^{3\alpha} p}{Zc^2 M}\right] \left[1 - \frac{ZGm(r^\alpha)}{c^2 r^\alpha}\right]^{-1} .  \tag{38}$$

When combining Equation (38) with the formula for the mass Equation (35) and a microscopic explanation for the relationship between pressure and energy density, this equation yields the equilibrium solution for pressure in a compact star. The relation between the energy density and pressure of the fluid given by the polytropic equation of state

$$\rho = \rho_c \theta^n, \tag{39}$$

$$p = K\rho^{1+\frac{1}{n}}, \tag{40}$$

where $n$ is the polytropic index and $K$ is the pressure constant that the thermal characteristics of a given fluid sphere must determine and $\theta$ is called the Emden function (the ratio of the density to the central density $\rho/\rho_c$).

Inserting Equations (39) and (40) in Equation (33) yields

$$\Gamma(\alpha+1)\frac{k\rho_c^{1+\frac{1}{n}}}{c^2}(1+n)D_r^\alpha \theta + \left(\frac{k\rho_c^{1+\frac{1}{n}}}{c^2}\theta + \rho_c\right)D_r^\alpha v = 0, \tag{41}$$

Now write Equation (41) as

$$\sigma \Gamma(\alpha+1)(1+n)D_r^\alpha \theta + (1+\sigma\,\theta)D_r^\alpha v = 0. \tag{42}$$

where the relativistic parameter $\sigma$ is given by

$$\sigma = \frac{k\rho_c^{1/n}}{c^2}.$$

From Equations (34, 36), we obtain

$$\frac{e^{-2\lambda}}{r^{2\alpha}}(1+Zr^\alpha D_r^\alpha v) - \frac{1}{r^{2\alpha}} = \frac{8\pi G}{c^4}p, \tag{43}$$

$$\frac{1}{r^{2\alpha}}\left(1-\frac{ZGm(r)}{r^\alpha c^2}\right)(1+Zr^\alpha D_r^\alpha v) - \frac{1}{r^{2\alpha}} = \frac{8\pi G}{c^4}p, \tag{44}$$

inserting Equations (39) and (40) in Equation (44) and put $r = a\xi$, we get

$$\xi^{2\alpha}D_\xi^\alpha \theta - \frac{\sigma(1+n)}{(1+\sigma\,\theta)}\frac{(1+\sigma\,\theta)}{\sigma(1+n)}\frac{AZG\,m}{c^2}\xi^\alpha D_\xi^\alpha \theta - \frac{(1+\sigma\,\theta)}{\sigma(1+n)}\frac{G\,m\,A}{c^2}$$
$$-\frac{(1+\sigma\,\theta)}{\sigma(1+n)}\frac{8\pi G}{A^2 Z c^4}\sigma\,\theta\,\rho\xi^{3\alpha} = 0 \tag{45}$$

where

$$A = \left(\frac{8\pi G\rho_c}{Z\sigma(n+1)c^2}\right)^{1/2},$$

$$\upsilon(\xi) = \frac{m}{M} = \frac{ZA^3 m}{8\pi\rho_c} = \frac{G\, m\, A}{\sigma(1+n)c^2}.$$

Rearrange terms; the FTOV equations have the form

$$\xi^{2\alpha} D_\xi^\alpha \theta - Z\sigma(n+1)\xi^\alpha \upsilon D_\xi^\alpha \theta + \upsilon + \upsilon\sigma\theta + \sigma\xi^\alpha \theta D_\xi^\alpha \upsilon + \sigma^2 \xi^\alpha \theta^2 D_\xi^\alpha \upsilon = 0, \tag{46}$$

and

$$D_\xi^\alpha \upsilon = \xi^2 \theta^n, \tag{47}$$

with the initial conditions

$$\theta(0) = 1 \quad , \upsilon(0) = 0 \quad , D_\xi^{\alpha\alpha} \theta = D_\xi^\alpha \left( D_\xi^\alpha \theta \right).$$

## 4. Series Solution of FTOV equation

### 4.1 Successive fractional derivatives of the Emden Function

The FTOV equations could be written as

$$\xi^{2\alpha} D_\xi^\alpha \theta - Z\sigma(n+1)\xi^\alpha \upsilon D_\xi^\alpha \theta + \upsilon + \upsilon\sigma\theta + \sigma\xi^\alpha \theta D_\xi^\alpha \upsilon + \sigma^2 \xi^\alpha \theta^2 D_\xi^\alpha \upsilon = 0, \tag{48}$$

and

$$D_\xi^\alpha \upsilon = \xi^{2\alpha} \theta^n, \tag{49}$$

subject to the initial condition

$$\theta(0) = 1 \quad , \upsilon(0) = 0 \quad , D_\xi^{\alpha\alpha} \theta = D_\xi^\alpha \left( D_\xi^\alpha \theta \right). \tag{50}$$

Considering a series expansion in the form

$$\theta(\xi) = \sum_{m=0}^{\infty} A_m \xi^{2m}, \tag{51}$$

we can rewrite Equation (51) in the form of fractional calculus as

$$\theta\left(\xi^{\alpha}\right) = A_0 + A_2\xi^{2\alpha} + A_4\xi^{4\alpha} + A_3\xi^{6\alpha} + \dots \quad (52)$$

where $\theta(0) = A_0 = 1$, then

$$\theta\left(\xi^{\alpha}\right) = 1 + \sum_{m=1}^{\infty} A_m \xi^{2\alpha m}, \quad (53)$$

applying Jumarie's mRL derivative, Equation (4), to Equation (53) yields

$$D_{\xi}^{\alpha}\theta = \sum_{m=1}^{\infty} \frac{\Gamma(2m+1)}{\Gamma(2m-\alpha+1)} A_m \xi^{2m-\alpha},$$

or

$$D_{\xi}^{\alpha}\theta = \sum_{m=1}^{\infty} U_m A_m \xi^{\alpha m-\alpha} = \sum_{m=0}^{\infty} U_{m+1} A_{m+1} \xi^{\alpha m+\alpha}, \quad (54)$$

where

$$U_m = \frac{\Gamma(2\alpha m+1)}{\Gamma(2\alpha m-\alpha+1)}, \quad (55)$$

and

$$U_{m+1} = \frac{\Gamma(2\alpha m+2\alpha+1)}{\Gamma(2\alpha m+\alpha+1)}. \quad (56)$$

So, Equation (54) could be written as

$$D_{\xi}^{\alpha}\theta = \frac{A_1\Gamma(2\alpha+1)}{\Gamma(\alpha+1)}\xi^{\alpha} + \frac{A_2\Gamma(4\alpha+1)}{\Gamma(3\alpha+1)}\xi^{3\alpha} + \frac{A_3\Gamma(6\alpha+1)}{\Gamma(5\alpha+1)}\xi^{5\alpha} + \frac{A_4\Gamma(8\alpha+1)}{\Gamma(7\alpha+1)}\xi^{7\alpha}\dots.$$

Performing the second derivative of $\theta$, gives

$$D_{\xi}^{\alpha\alpha}\theta = \frac{A_1\Gamma(2\alpha+1)}{\Gamma(\alpha+1)}D_{\xi}^{\alpha}\xi^{\alpha} + \frac{A_2\Gamma(4\alpha+1)}{\Gamma(3\alpha+1)}D_{\xi}^{\alpha}\xi^{3\alpha} + \frac{A_3\Gamma(6\alpha+1)}{\Gamma(5\alpha+1)}D_{\xi}^{\alpha}\xi^{5\alpha} + \frac{A_4\Gamma(8\alpha+1)}{\Gamma(7\alpha+1)}D_{\xi}^{\alpha}\xi^{7\alpha}\dots,$$

gives

$$D_\xi^{\alpha\alpha}\theta = \frac{A_1\Gamma(2\alpha+1)}{\Gamma(1)} + \frac{A_2\Gamma(4\alpha+1)}{\Gamma(2\alpha+1)}\xi^{2\alpha} + \frac{A_3\Gamma(6\alpha+1)}{\Gamma(4\alpha+1)}\xi^{4\alpha} + \frac{A_4\Gamma(8\alpha+1)}{\Gamma(6\alpha+1)}\xi^{4\alpha}\ldots \quad . \tag{57}$$

at $\xi^\alpha = 0$, Equation (57) becomes

$$D_\xi^{\alpha\alpha}\theta(0) = A_1\Gamma(2\alpha+1). \tag{58}$$

Differentiate Equation (58) $2j$ times

$$D_\xi^{\overbrace{\alpha\ldots\alpha}^{2j\,times}}\theta(0) = A_j\Gamma(2j\alpha+1) \quad , \tag{59}$$

where $A_j$ are constants to be determined.

## 4.2 Fractional Derivative of the Emden Function Raised to Powers

Taking the fractional derivative of the product of two functions $\theta^2$, which corresponds to $\theta$ times $\theta$, yields the fractional derivative of the Emden function $\theta^n$. The Emden function raised to powers could be written as

$$\theta^n(\xi) = \sum_{m=0}^\infty Q_m \xi^{2m} = G(\xi), \tag{60}$$

Write Equation (60) in the fractional form as

$$\theta^n(\xi^\alpha) = \sum_{m=0}^\infty Q_m \xi^{2\alpha m} = G(\xi^\alpha), \tag{61}$$

following the same manipulation of [18], we get the coefficient $Q_m$

$$Q_m = \frac{1}{\Gamma(m\alpha+1)A_0}\sum_{i=1}^m \frac{(m-1)!\Gamma(\alpha(m-i)+1)\Gamma(2i\alpha+1)}{i!(m-i)!}(in-m+i)A_i Q_{m-i} \quad \forall m\geq 1. \tag{62}$$

## 4.3 Fractional Derivative of the Series Expansion of the Relativistic Function

Multiply Equation (54) with $\xi^{2\alpha}$, and we get

$$\xi^{2\alpha} D_\xi^\alpha \theta = \sum_{m=0}^{\infty} U_{m+1} A_{m+1} \xi^{2\alpha m+3\alpha}, \tag{63}$$

Now using the series expansion of $\theta^n$, Equation (61), the relativistic function (Equation (49)) writes

$$D_\xi^\alpha \upsilon = \xi^{2\alpha} \theta^n = \sum_{m=0}^{\infty} Q_m \xi^{2\alpha m+2\alpha}, \tag{64}$$

integrating the last equation gives

$$\upsilon = \sum_{m=0}^{\infty} \frac{\Gamma(2\alpha m+2\alpha+1)}{\Gamma(2\alpha m+3\alpha+1)} Q_m \xi^{2\alpha m+3\alpha},$$

or

$$\upsilon = \sum_{m=0}^{\infty} N_m \xi^{2\alpha m+3\alpha}, \tag{65}$$

where

$$N_m = \frac{\Gamma(2\alpha m+2\alpha+1)}{\Gamma(2\alpha m+3\alpha+1)} Q_m. \tag{66}$$

Combining Equations (55) and (56), we get

$$\xi^\alpha \upsilon D_\xi^\alpha \theta = \left( \sum_{m=0}^{\infty} U_{m+1} A_{m+1} \xi^{2\alpha m+2\alpha} \right) \left( \sum_{m=0}^{\infty} N_m \xi^{2\alpha m+3\alpha} \right), \tag{67}$$

if $f_m = A_{m+1} U_{m+1}$, then

$$\xi^\alpha \upsilon D_\xi^\alpha \theta = \left( \sum_{m=0}^{\infty} f_m \xi^{2\alpha m+3\alpha} \right) \left( \sum_{m=0}^{\infty} N_m \xi^{2\alpha m+3\alpha} \right), \tag{68}$$

where

$$f_m = f_0 N_m + f_1 N_{m-1} + \ldots + f_m N_0,$$

$$\xi^\alpha \upsilon D_\xi^\alpha \theta = \sum_{m=0}^{\infty} \gamma_m \xi^{2\alpha m+3\alpha}, \tag{69}$$

where:

$$\gamma_m = \sum_{i=0}^{m} f_i N_{m-i},$$
$$f_i = A_{i+1} U_{i+1}.$$
(70)

Multiply Equation (65) by Equation (51), and we get:

$$\upsilon\theta = \left(\sum_{m=0}^{\infty} N_m \xi^{2\alpha m+3\alpha}\right)\left(\sum_{m=0}^{\infty} A_{2m} \xi^{2\alpha m}\right),$$

$$\upsilon\theta = \sum_{m=0}^{\infty} \eta_m \xi^{2\alpha m+3\alpha},$$
(71)

where

$$\eta_m = \sum_{i=0}^{m} A_i N_{m-i}.$$

Using Equation (51) and Equation (64), we can write

$$\xi^\alpha \theta D_\xi^\alpha \upsilon = \left(\sum_{m=0}^{\infty} Q_m \xi^{2\alpha m+2\alpha}\right)\left(\sum_{m=0}^{\infty} A_m \xi^{2\alpha m}\right)\xi^\alpha,$$

$$\xi^\alpha \theta D_\xi^\alpha \upsilon = \left(\sum_{m=0}^{\infty} Q_m \xi^{2\alpha m+3\alpha}\right)\left(\sum_{m=0}^{\infty} A_m \xi^{2\alpha m}\right),$$

$$\xi^\alpha \theta D_\xi^\alpha \upsilon = \sum_{m=0}^{\infty} \beta_m \xi^{2\alpha m+3\alpha},$$
(72)

where

$$\beta_m = \sum_{i=0}^{m} A_i Q_{m-i}.$$

Multiply Equation (72) by $\theta$ yields

$$\xi^\alpha \theta^2 D_\xi^\alpha \upsilon = \left(\sum_{m=0}^{\infty} \beta_m \xi^{2\alpha m+3\alpha}\right)\left(\sum_{m=0}^{\infty} A_m \xi^{2\alpha m}\right),$$

$$\xi^\alpha \theta^2 D_\xi^\alpha \upsilon = \sum_{m=0}^{\infty} \zeta_m \xi^{2\alpha m+3\alpha},$$
(73)

where

$$\zeta_m = \sum_{i=0}^{m} A_i \beta_{m-i}.$$

Inserting Equations (29), (31), (35), (37), (39), and (39) into Equation (14) gives

$$\sum_{m=0}^{\infty} U_{m+1} A_{m+1} \xi^{2\alpha m+3\alpha} - Z\sigma(n+1) \sum_{m=0}^{\infty} \gamma_m \xi^{2\alpha m+3\alpha} + \sum_{m=0}^{\infty} N_m \xi^{2\alpha m+3\alpha}$$
$$+ \sigma \sum_{m=0}^{\infty} \eta_m \xi^{2\alpha m+3\alpha} + \sigma \sum_{m=0}^{\infty} \beta_m \xi^{2\alpha m+3\alpha} + \sigma^2 \sum_{m=0}^{\infty} \xi_m \xi^{2\alpha m+3\alpha} = 0.$$

We can rewrite this equation in the like power of $\xi$; we have

$$U_{m+1} A_{m+1} - Z\sigma(n+1) \gamma_{m-1} + N_m + \sigma \eta_m + \sigma \beta_m + \sigma^2 \xi_m = 0, \tag{74}$$

where

$$\gamma_{m-1} = \sum_{i=0}^{m} f_i N_{m-i-1}.$$

## 4.4 The Recurrence Relations

The coefficients of the series expansion can be calculated using the two recurrence relations

$$A_{m+1} = \frac{\sigma}{U_{m+1}} \left( Z(n+1) \gamma_{m-1} - \eta_m - \beta_m - \sigma^2 \xi_m \right) - \frac{N_m}{U_{m+1}}, \tag{75}$$

and

$$Q_m = \frac{1}{\Gamma(m\alpha+1) A_0} \sum_{i=1}^{m} \frac{(m-1)! \Gamma(\alpha(m-i)+1) \Gamma(i\alpha+1)}{i!(m-i)!} (in-m+i) A_i Q_{m-i} \quad \forall m \geq 1, \tag{76}$$

where

$$\zeta_m = \sum_{i=0}^{m} A_i \beta_{m-i}, \quad \beta_m = \sum_{i=0}^{m} A_i Q_{m-i}, \quad \eta_m = \sum_{i=0}^{m} A_i N_{m-i}, \quad \gamma_{m-1} = \sum_{i=0}^{m} f_i N_{m-i}, \quad f_i = A_{i+1} U_{i+1},$$
$$N_m = \frac{\Gamma(2\alpha m + 2\alpha + 1)}{\Gamma(2\alpha m + 3\alpha + 1)} Q_{2m}, \quad U_m = \frac{\Gamma(2\alpha m + 1)}{\Gamma(2\alpha m - \alpha + 1)}.$$

Applying the initial condition in Equation (75) to $m=0$, we obtain

$$A_1 = \frac{\sigma}{U_1}\left(-\eta_0 - \beta_0 - \sigma^2 \xi_0\right) - \frac{N_0}{U_1}, \tag{77}$$

If we put $\alpha = 1$, we have the solution to the integer TOV equation

$$A_1 = -\frac{1}{6} - \frac{\sigma}{2}\left(\frac{4}{3} + \sigma\right), \tag{78}$$

set $\sigma = 0$ in the last equation, we get the first series term of the solution of the Newtonian LE as

$$A_1 = -\frac{1}{6}. \tag{79}$$

## 5. Numerical Results

To compute the Emden function ($\theta$, Equation (53)) and the relativistic function ($\upsilon$, Equation (64)), we used the two recurrence relations represented by Equations (75) and (76). We developed a MATHEMATICA code to calculate some relativistic polytropic models (i.e., mass-radius relation, density, pressure, and temperature profiles) for the range of the polytropic index n=0-3 subject to all possible values of the relativistic parameters. We run the code to calculate the Emden function at the polytropic index n=1.5. Figure (1) shows the results for the Emden function, where we can assess the divergent behavior of the series as found for the Newtonian LE equation; the series converges for $\xi \leq 2.5$ and then diverges (see, for instance, Nouh 2004).

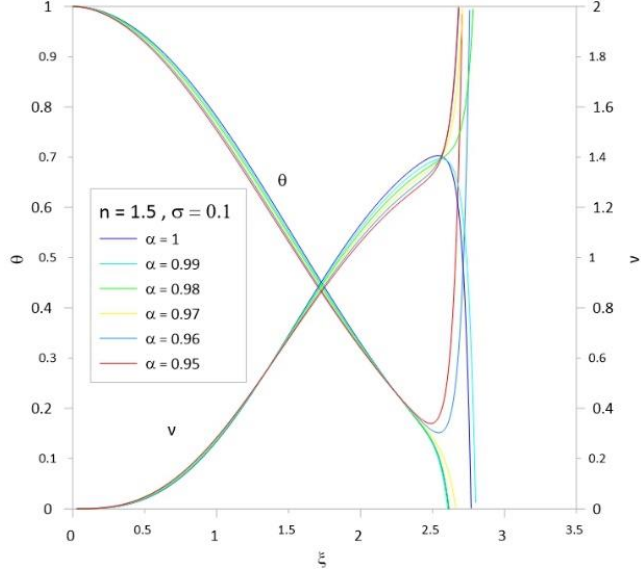

Figure 1: The Emden and relativistic function computed by series expansion without acceleration. The models were calculated for n=1.5, σ=0.1. It is well seen that the series diverges for all values of $\alpha$ at $\xi > 2.5$.

To accelerate the convergence of the series, we use the scheme developed by [10,12] by combining the two techniques for Euler-Abel transformation and Padé approximation. Following the above procedure of calculations, we obtained the results of Tables (1-6); column 1 is the relativistic parameter $\sigma$, column 2 is the first zero of the numerical solution $(\xi_{1N})$ for the integer case ($\alpha = 1$) obtained by [10,29], and columns 3 to 8 is the first zero of an analytical solution for TOV function at different values of the fractional parameter $(\alpha = 1, 0.99, 0.98, 0.97, 0.96, 0.95)$. In Tables (7-12), we listed the mass function calculated at the first zero $(\upsilon(\xi_1))$, where column 1 is the relativistic parameter $\sigma$, column 2 is $\upsilon(\xi_1)$ calculated for the integer case ($\alpha = 1$), and columns 3 to 8 are $\upsilon(\xi_1)$ calculated at different values of the fractional parameter $(\alpha = 1, 0.99, 0.98, 0.97, 0.96, 0.95)$. The calculation for each fractional model is stopped when the relativistic parameter is equal to the maximum value $\sigma_{max} = n/n+1$ (where the sound velocity must be smaller than the speed of light). Figures (2-7) display the distribution of the Emden and the mass functions computed for different fractional parameters and polytropic indices range $n = 0-3$.

Table 1. The radius of convergence of the fractional polytrope with $n = 0$.

| | $\xi_1$ | | | | | |
|---|---|---|---|---|---|---|
| $\sigma$ | $\alpha=1$ | $\alpha=0.99$ | $\alpha=0.98$ | $\alpha=0.97$ | $\alpha=0.96$ | $\alpha=0.95$ |
| 0 | 2.449 | 2.421 | 2.393 | 2.366 | 2.340 | 2.314 |

Table 2. The radius of convergence of the fractional polytrope with $n = 0.5$.

| | $\xi_1$ | | | | | |
|---|---|---|---|---|---|---|
| $\sigma$ | $\alpha=1$ | $\alpha=0.99$ | $\alpha=0.98$ | $\alpha=0.97$ | $\alpha=0.96$ | $\alpha=0.95$ |
| 0 | 2.752 | 2.729 | 2.708 | 2.688 | 2.669 | 2.652 |
| 0.1 | 2.289 | 2.279 | 2.271 | 2.263 | 2.261 | 2.257 |
| 0.2 | 2.000 | 1.997 | 1.996 | 1.996 | 1.998 | 2.003 |
| 0.3 | 1.801 | 1.802 | 1.804 | 1.809 | 1.816 | 1.835 |

Table 3. The radius of convergence of the fractional polytrope with $n = 1$.

| | $\xi_1$ | | | | | |
|---|---|---|---|---|---|---|
| $\sigma$ | $\alpha=1$ | $\alpha=0.99$ | $\alpha=0.98$ | $\alpha=0.97$ | $\alpha=0.96$ | $\alpha=0.95$ |
| 0 | 3.141 | 3.124 | 3.108 | 3.093 | 3.080 | 3.068 |
| 0.1 | 2.599 | 2.597 | 2.597 | 2.596 | 2.597 | 2.599 |
| 0.2 | 2.278 | 2.286 | 2.288 | 2.296 | 2.306 | 2.315 |
| 0.3 | 2.064 | 2.077 | 2.084 | 2.095 | 2.112 | 2.124 |
| 0.4 | 1.912 | 1.926 | 1.938 | 1.954 | 1.971 | 1.989 |
| 0.5 | 1.801 | 1.816 | 1.833 | 1.850 | 1.870 | 1.891 |

Table 4. The radius of convergence of the fractional polytrope with $n = 1.5$.

| | $\xi_1$ | | | | | |
|---|---|---|---|---|---|---|
| $\sigma$ | $\alpha=1$ | $\alpha=0.99$ | $\alpha=0.98$ | $\alpha=0.97$ | $\alpha=0.96$ | $\alpha=0.95$ |
| 0 | 3.653 | 3.642 | 3.632 | 3.623 | 3.615 | 3.608 |
| 0.1 | 3.046 | 3.047 | 3.055 | 3.065 | 3.077 | 3.077 |
| 0.2 | 2.696 | 2.710 | 2.725 | 2.741 | 2.758 | 2.776 |
| 0.3 | 2.496 | 2.514 | 2.533 | 2.533 | 2.580 | 2.604 |
| 0.4 | 2.363 | 2.414 | 2.399 | 2.392 | 2.508 | 2.479 |
| 0.5 | 2.270 | 2.286 | 2.297 | 2.354 | 2.403 | 2.349 |
| 0.6 | 2.219 | 2.240 | 2.264 | 2.308 | 2.349 | 2.399 |

Table 5. The radius of convergence of the fractional polytrope with $n = 2$.

| | $\xi_1$ | | | | | |
|---|---|---|---|---|---|---|
| $\sigma$ | $\alpha=1$ | $\alpha=0.99$ | $\alpha=0.98$ | $\alpha=0.97$ | $\alpha=0.96$ | $\alpha=0.95$ |
| 0 | 4.352 | 4.345 | 4.338 | 4.330 | 4.320 | 4.310 |
| 0.1 | 3.698 | 3.691 | 3.713 | 3.751 | 3.763 | 3.788 |
| 0.2 | 3.398 | 3.425 | 3.453 | 3.481 | 3.509 | 3.538 |
| 0.3 | 3.271 | 3.311 | 3.347 | 3.386 | 3.426 | 3.465 |
| 0.4 | 3.245 | 3.297 | 3.349 | 3.404 | 3.457 | 3.513 |
| 0.5 | 3.291 | 3.364 | 3.438 | 3.513 | 3.588 | 3.666 |
| 0.6 | 3.398 | 3.496 | 3.585 | 3.699 | 3.792 | 3.912 |
| 0.66667 | 3.490 | 3.608 | 3.790 | 3.853 | 3.985 | 4.150 |

Table 6. The radius of convergence of the fractional polytrope with $n = 3$.

| | $\xi_1$ | | | | | |
|---|---|---|---|---|---|---|
| $\sigma$ | $\alpha=1$ | $\alpha=0.99$ | $\alpha=0.98$ | $\alpha=0.97$ | $\alpha=0.96$ | $\alpha=0.95$ |
| 0 | 6.894 | 6.658 | 6.464 | 6.309 | 6.182 | 6.075 |
| 0.001 | 6.864 | 6.887 | 6.863 | 6.929 | 6.943 | 6.909 |
| 0.003 | 6.850 | 6.868 | 6.893 | 6.917 | 6.952 | 6.976 |
| 0.005 | 6.837 | 6.857 | 6.884 | 6.904 | 6.934 | 6.973 |
| 0.007 | 6,835 | 6.564 | 6.874 | 6.891 | 6.920 | 6.950 |
| 0.009 | 6.833 | 6.855 | 6.877 | 6.880 | 6.929 | 6.954 |
| 0.01 | 6.821 | 6.831 | 6.858 | 6.878 | 6.898 | 6.931 |
| 0.03 | 6.707 | 6.772 | 6.813 | 6.819 | 6.813 | 6.864 |
| 0.05 | 6.710 | 6.650 | 6.560 | 6.766 | 6.784 | 6.851 |
| 0.07 | 6.723 | 6.938 | 7.001 | 7.112 | 7.162 | 7.195 |
| 0.09 | 6.811 | 6.963 | 7.065 | 7.119 | 7.237 | 7.295 |
| 0.1 | 6.872 | 7.097 | 7.107 | 7.031 | 7.119 | 7.452 |
| 0.2 | 7.941 | 8.039 | 7.905 | 7.294 | 7.063 | 7.876 |
| 0.3 | 10.841 | 10.342 | 10.937 | 10.549 | 10.801 | 10.633 |

Table 7. Mass Function of the fractional polytrope with $n = 0$.

| | $\upsilon(\xi_1)$ | | | | | |
|---|---|---|---|---|---|---|
| $\sigma$ | $\alpha=1$ | $\alpha=0.99$ | $\alpha=0.98$ | $\alpha=0.97$ | $\alpha=0.96$ | $\alpha=0.95$ |
| 0 | 4.8960 | 4.6954 | 4.5038 | 4.3263 | 4.1617 | 4.0043 |

Table 8. Mass Function of the fractional polytrope with $n = 0.5$.

| | $\upsilon(\xi_1)$ | | | | | |
|---|---|---|---|---|---|---|
| $\sigma$ | $\alpha=1$ | $\alpha=0.99$ | $\alpha=0.98$ | $\alpha=0.97$ | $\alpha=0.96$ | $\alpha=0.95$ |
| 0 | 3.7909 | 3.6986 | 3.5987 | 3.4945 | 3.3888 | 3.2848 |
| 0.1 | 2.2203 | 2.1714 | 2.1235 | 2.0748 | 2.0305 | 2.0174 |
| 0.2 | 1.4738 | 1.4483 | 1.4235 | 1.3984 | 1.3733 | 1.3480 |
| 0.3 | 1.0613 | 1.0471 | 1.0324 | 1.0179 | 1.0022 | 0.9863 |

Table 9. Mass Function of the fractional polytrope with $n = 1$.

| | $\upsilon(\xi_1)$ | | | | | |
|---|---|---|---|---|---|---|
| $\sigma$ | $\alpha=1$ | $\alpha=0.99$ | $\alpha=0.98$ | $\alpha=0.97$ | $\alpha=0.96$ | $\alpha=0.95$ |
| 0 | 3.1416 | 3.0451 | 2.9528 | 2.8643 | 2.7796 | 2.6985 |
| 0.1 | 1.7514 | 1.7167 | 1.6831 | 1.6505 | 1.6188 | 1.5881 |
| 0.2 | 1.1426 | 1.1275 | 1.1127 | 1.0983 | 1.0842 | 1.0704 |
| 0.4 | 0.8192 | 0.8119 | 0.8049 | 0.7980 | 0.7911 | 0.7846 |
| 0.5 | 0.6249 | 0.6215 | 0.6182 | 0.6149 | 0.6116 | 0.6085 |
| 0.6 | 0.4984 | 0.4969 | 0.4956 | 0.4940 | 0.4928 | 0.4917 |

Table 10. Mass Function of the fractional polytrope with $n = 1.5$.

| $\upsilon(\xi_1)$ |
|---|

| σ | α =1 | α = 0.99 | α = 0.98 | α = 0.97 | α = 0.96 | α = 0.95 |
|---|---|---|---|---|---|---|
| 0 | 2.7137 | 2.6668 | 2.6211 | 2.5769 | 2.5343 | 2.4941 |
| 0.1 | 1.4825 | 1.4717 | 1.4640 | 1.4560 | 1.4488 | 1.4405 |
| 0.2 | 0.9642 | 0.9704 | 0.9626 | 0.9597 | 0.9589 | 0.9593 |
| 0.3 | 0.6863 | 0.6888 | 0.6924 | 0.6966 | 0.7030 | 0.7134 |
| 0.4 | 0.5249 | 0.5299 | 0.5367 | 0.5412 | 0.5503 | 0.5554 |
| 0.5 | 0.4204 | 0.4260 | 0.4316 | 0.4380 | 0.4447 | 0.4488 |
| 0.6 | 0.3490 | 0.3549 | 0.3615 | 0.3737 | 0.3656 | 0.3829 |

Table 11. Mass Function of the fractional polytrope with $n = 2$.

| | $v(\xi_1)$ | | | | | |
|---|---|---|---|---|---|---|
| σ | α =1 | α = 0.99 | α = 0.98 | α = 0.97 | α = 0.96 | α = 0.95 |
| 0 | 2.4118 | 2.3918 | 2.3728 | 2.3937 | 2.3553 | 2.3340 |
| 0.1 | 1.2957 | 1.3084 | 1.3166 | 1.3260 | 1.3348 | 1.3477 |
| 0.2 | 0.8348 | 0.8477 | 0.8611 | 0.8745 | 0.8875 | 0.9003 |
| 0.3 | 0.6039 | 0.6189 | 0.6342 | 0.6493 | 0.6657 | 0.6811 |
| 0.4 | 0.4642 | 0.4784 | 0.4927 | 0.5078 | 0.5234 | 0.5414 |
| 0.5 | 0.3782 | 0.3916 | 0.4071 | 0.4210 | 0.4352 | 0.4491 |
| 0.6 | 0.3265 | 0.3343 | 0.3591 | 0.3552 | 0.3868 | 0.3979 |
| 0.66667 | 0.2815 | 0.3224 | 0.3584 | 0.3771 | 0.4304 | 0.4274 |

Table 12. Mass Function of the fractional polytrope with $n = 3$.

| | $v(\xi_1)$ | | | | | |
|---|---|---|---|---|---|---|
| σ | α =1 | α = 0.99 | α = 0.98 | α = 0.97 | α = 0.96 | α = 0.95 |
| 0 | 2.02059 | 2.111623 | 2.18202 | 2.2219 | 2.21187 | 2.24302 |
| 0.001 | 2.01684 | 2.11124 | 2.15300 | 2.20077 | 2.23836 | 2.26399 |
| 0.003 | 1.99952 | 2.08854 | 2.13234 | 2.17451 | 2.21357 | 2.24677 |
| 0.005 | 1.97217 | 2.07092 | 2.1120 | 2.15108 | 2.18773 | 2.22179 |
| 0.007 | 1.942892 | 2.02804 | 2.09632 | 2.13112 | 2.16509 | 2.19614 |
| 0.009 | 1.91368 | 2.06725 | 2.08976 | 2.11583 | 2.14891 | 2.1769 |
| 0.01 | 1.89885 | 2.07538 | 2.08744 | 2.11088 | 2.1370 | 2.16455 |
| 0.03 | 1.6245 | 1.7009 | 1.75646 | 1.80284 | 1.8439 | 1.88557 |

| 0.05 | 1.42668 | 1.49064 | 1.55204 | 1.60963 | 1.65237 | 1.6961 |
| 0.07 | 1.26794 | 1.33857 | 1.398787 | 1.4582 | 1.51095 | 1.55853 |
| 0.09 | 1.14184 | 1.20855 | 1.26883 | 1.32488 | 1.38185 | 1.43221 |
| 0.1 | 1.07302 | 1.1182 | 1.16447 | 1.19663 | 1.23351 | 1.27312 |
| 0.2 | 0.714168 | 0.760164 | 0.807264 | 0.84647 | 0.886431 | 0.961846 |
| 0.3 | 0.543307 | 0.596392 | 0.641067 | 0.707563 | 0.720947 | 0.804666 |

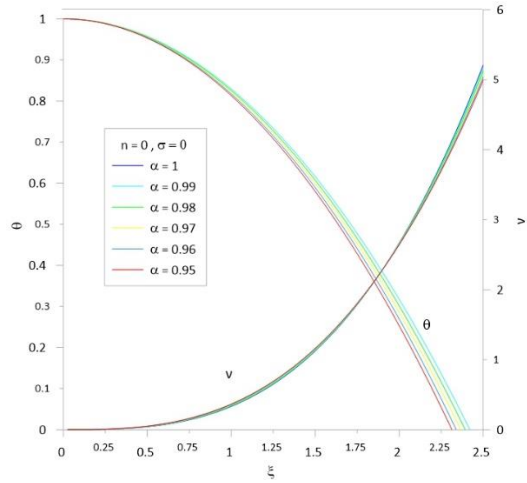

Figure 2: The Emden and relativistic functions for the relativistic fractional polytrope with n=0.

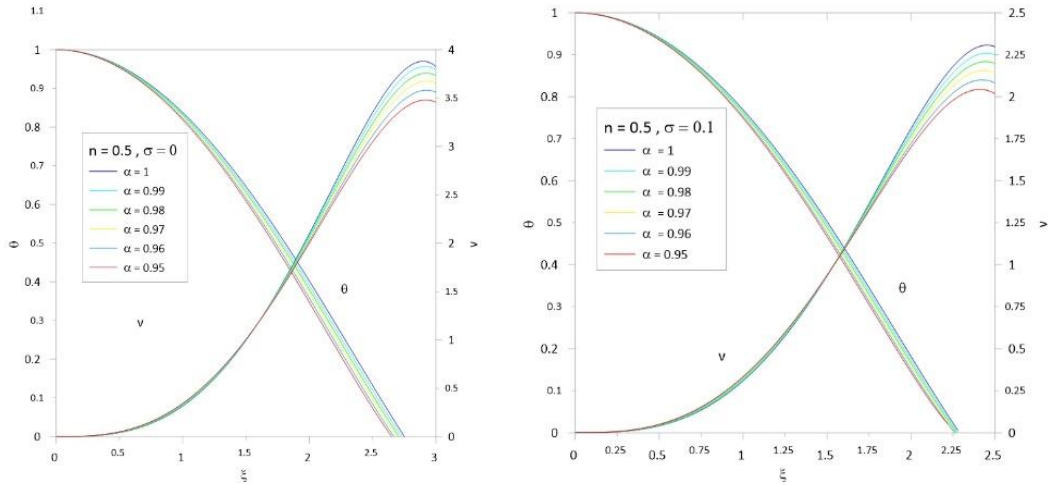

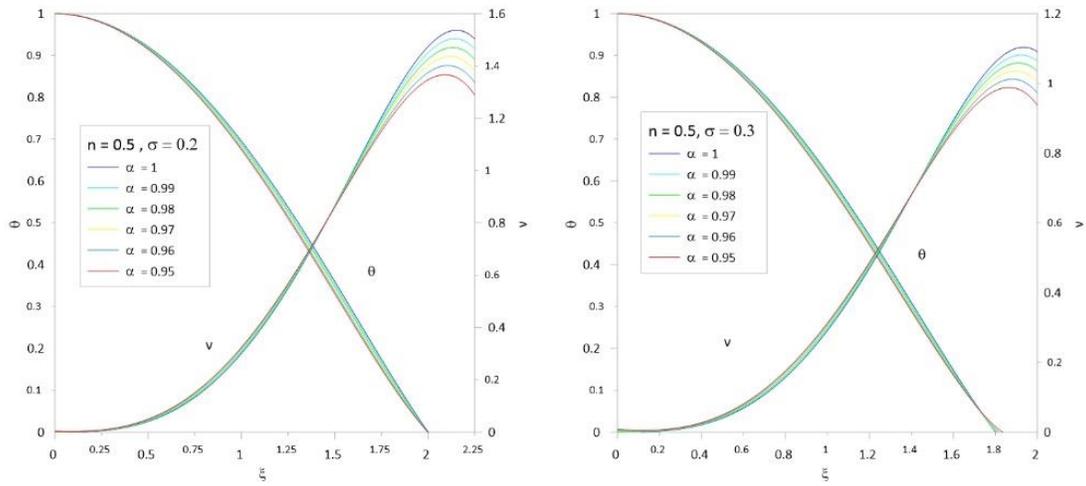

Figure 3: The Emden and relativistic functions for the relativistic fractional polytrope with n=0.5.

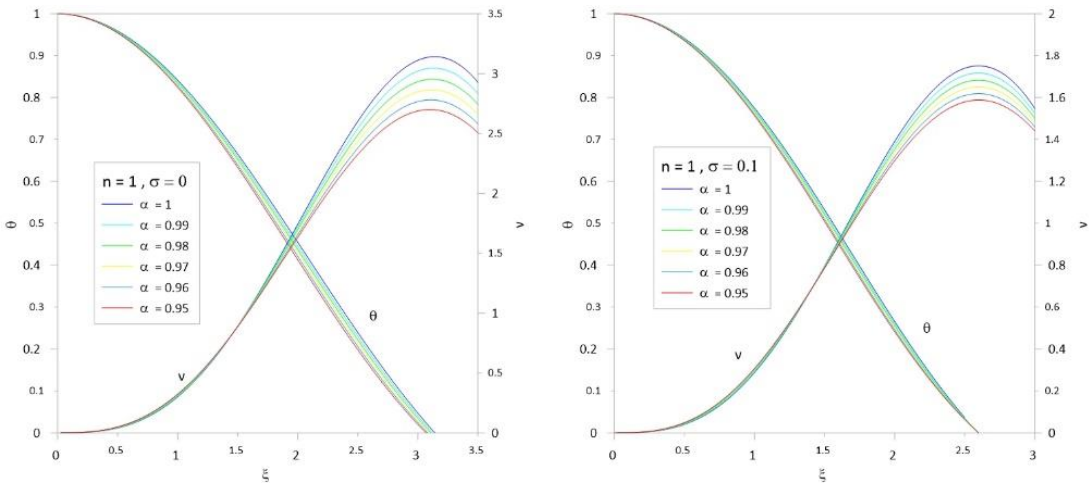

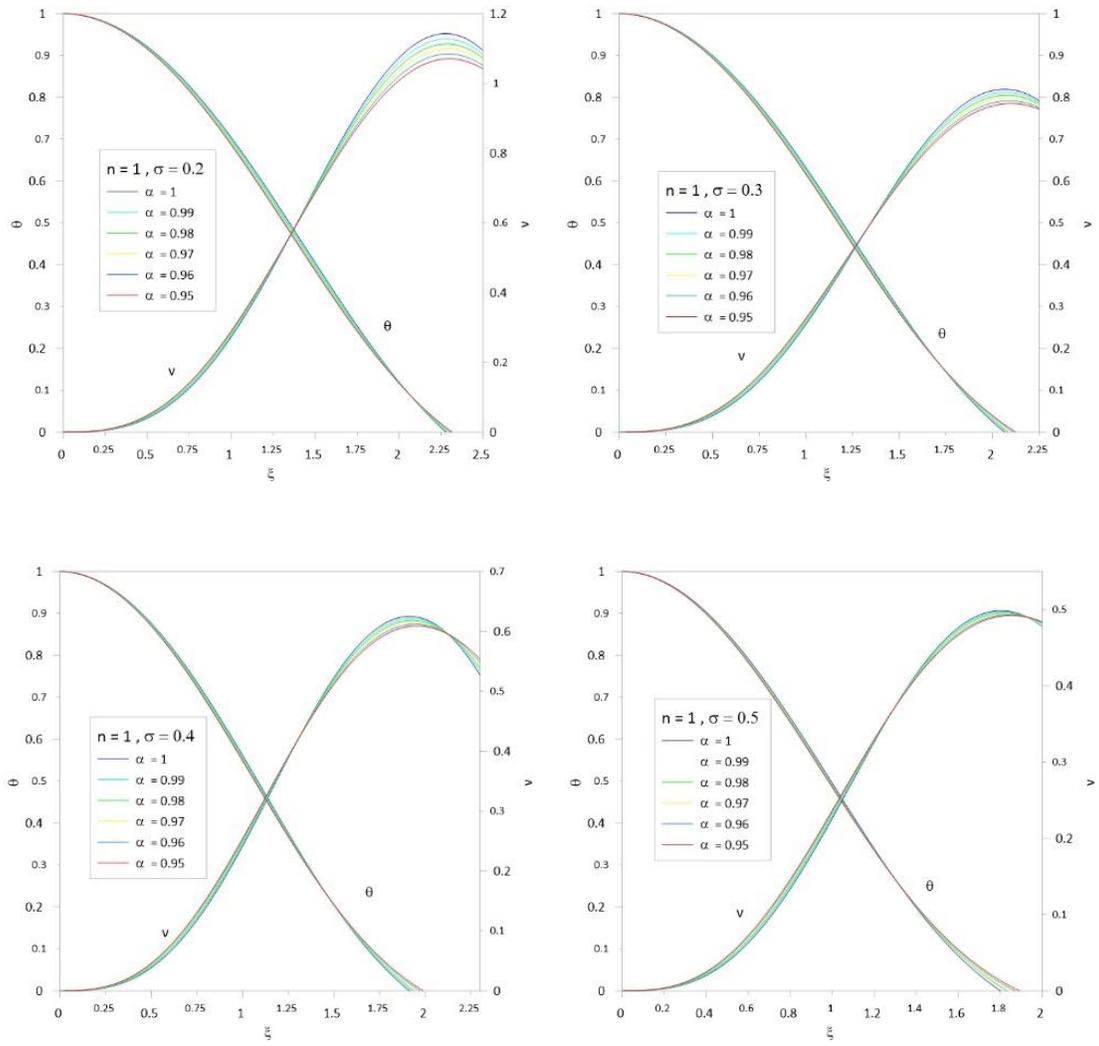

Figure 4: The Emden and relativistic functions for the relativistic fractional polytrope with n=1.

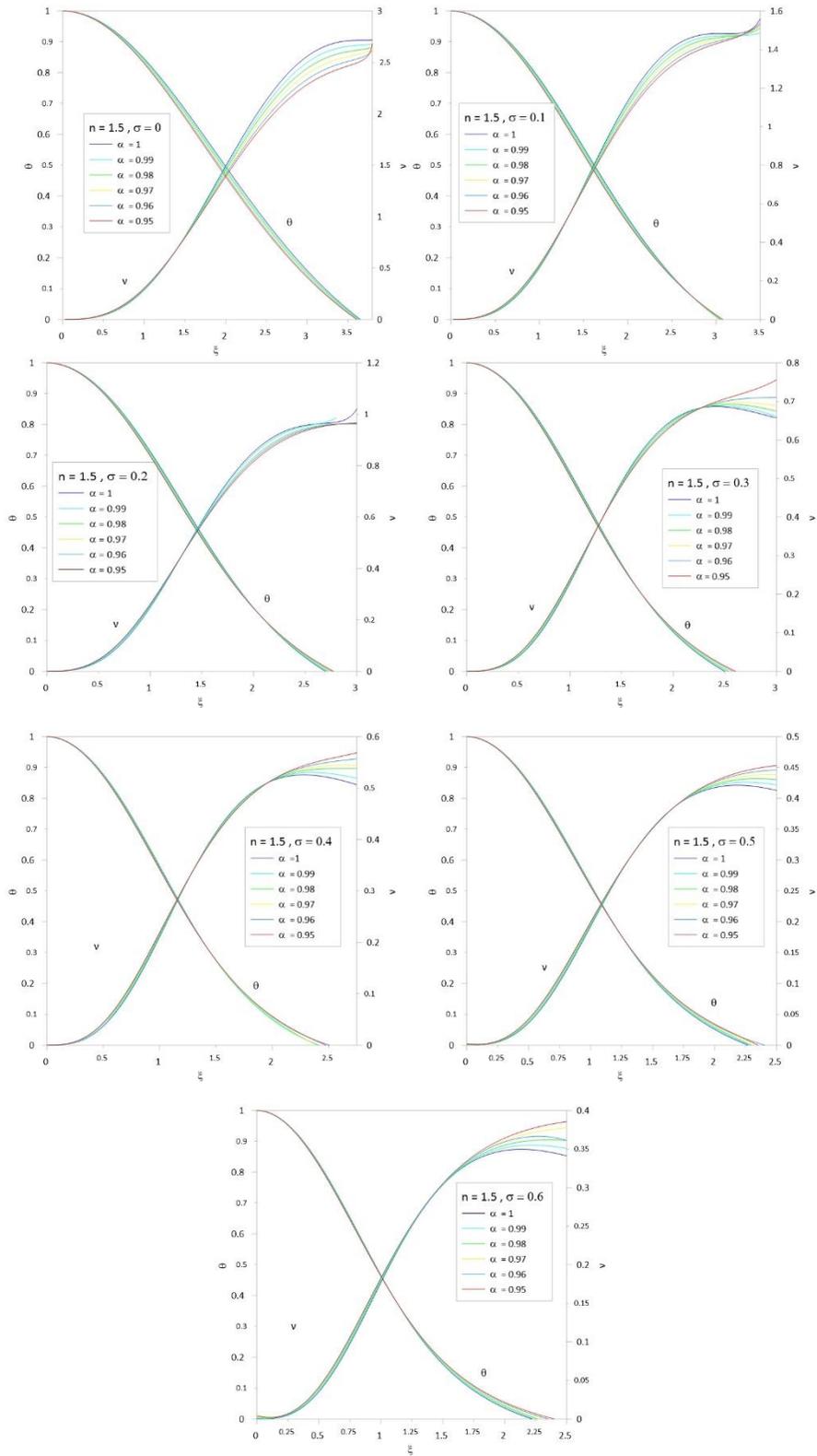

Figure 5: The Emden and relativistic functions for the relativistic fractional polytrope with n=1.5.

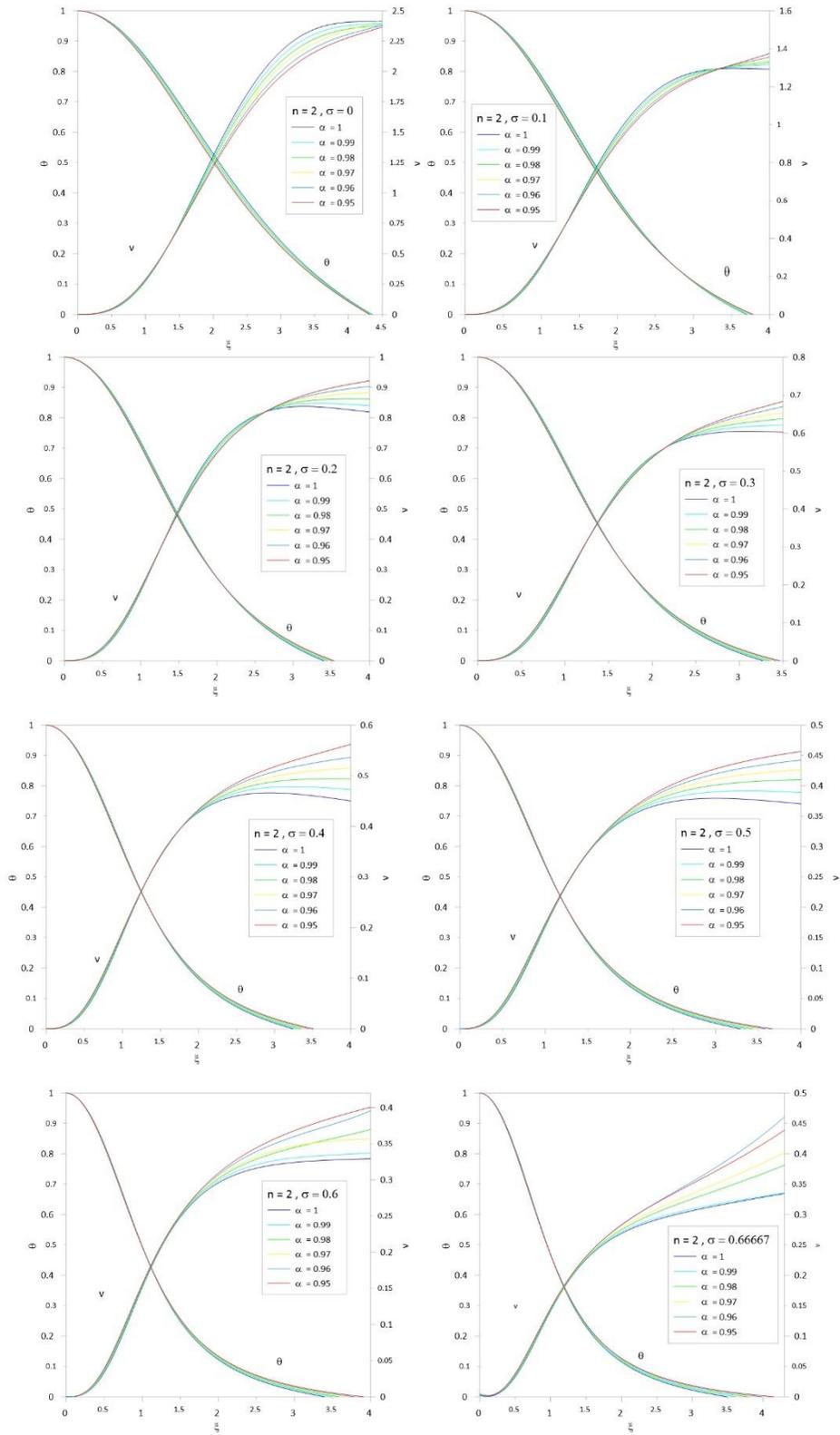

Figure 6: The Emden and relativistic functions for the relativistic fractional polytrope with n=2.

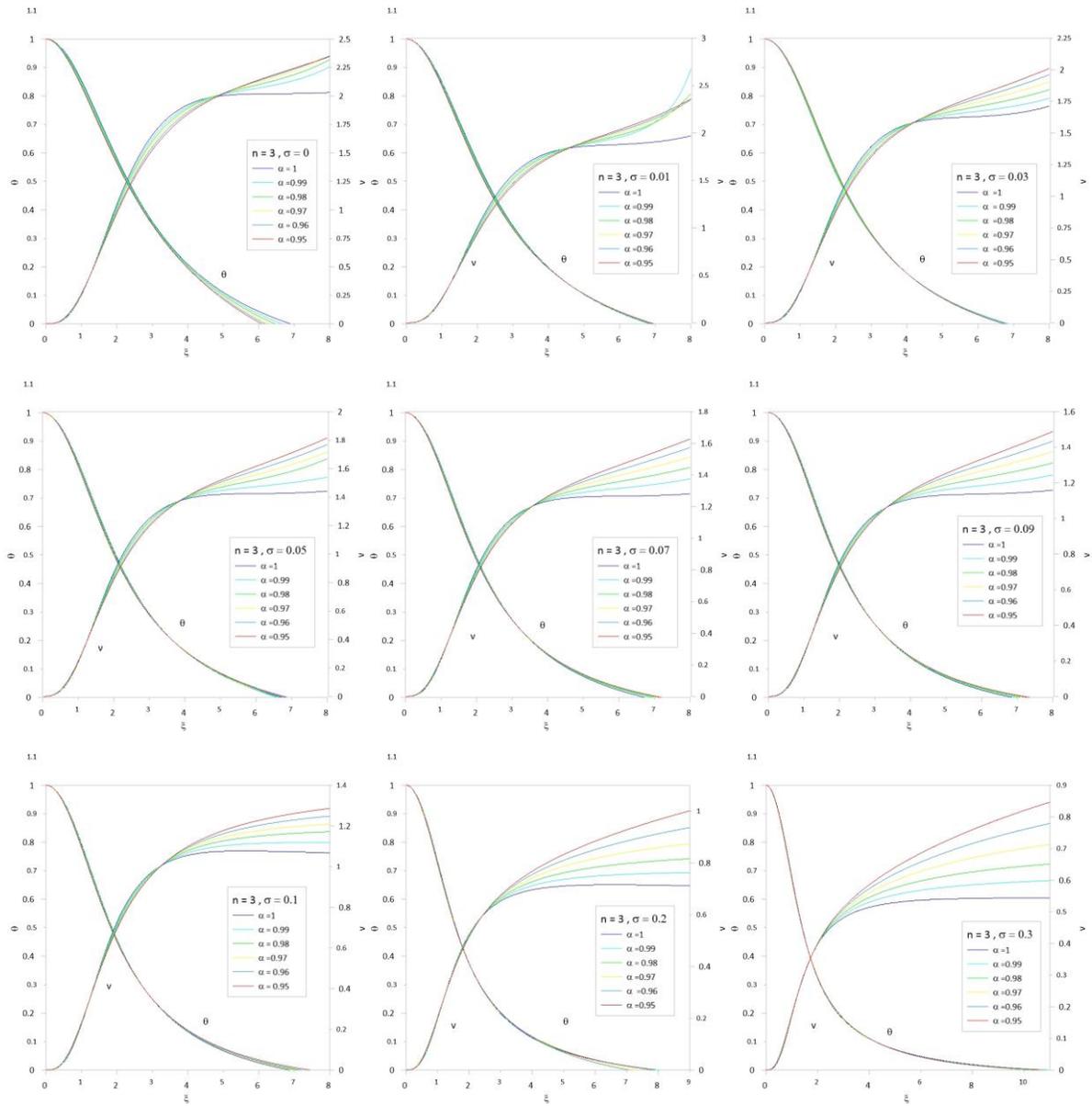

Figure 7: The Emden and relativistic functions for the relativistic fractional polytrope with n=3.

One can determine the radius $R$ and the mass $M$ of the polytrope from the following equations

$$R = A^{-1}\xi_1, \tag{80}$$

$$M = \frac{4\pi \rho_c}{A^3} v(\xi_1) = \left[\frac{1}{4\pi}\left(\frac{(n+1)c^2}{G}\right)^3 \left(\frac{K}{c^2}\right)^n\right]^{1/2} \tilde{M}, \tag{81}$$

where

$$\tilde{M} \equiv \sigma^{(3-n)/2}\, v(\xi_1),$$

$$A = \left(\frac{4\pi G \rho_c}{\sigma(n+1)c^2}\right)^{1/2}.$$

Figures (8-9) plot the ratios $\xi_{1f}/\xi_{1i}$ and $\upsilon_f(\xi_1)/\upsilon_i(\xi_1)$ (equivalent to the radius and mass ratios), where $\xi_{1i}$ and $\xi_{1f}$ are the zeroth of the integer and fractional Emden functions $\upsilon_i(\xi_1)$ and $\upsilon_f(\xi_1)$ are the integer and fractional mass functions; versus the relativistic parameter for different polytropic indices. The relativistic parameter $\sigma$ tends to increase the volume of the sphere [29], whereas the fractional parameter $\alpha$ tends to decrease the volume of the sphere [30]. The result of the two effects for n=0.5 is that the sphere's volume decreases with increasing $\sigma$ and increasing $\alpha$. For n=1, for all values of $\alpha$, the volume decreases when $\sigma$=0.1 and then increases when $\sigma$=0.2 and 0.3. For n=1.5 and n=2, the result of the two effects is opposite to the case of n=0.5 and n=1; the volume of the sphere decreases with increasing both $\sigma$ and $\alpha$. The mass ratio acts in nearly the same behavior as the radius ratio.

The compactification factor (CF) could be computed from the ratio of the mass function to the zero of the Emden function ($\upsilon(\xi_1)/\xi_1$). Figure (10) displays the variation of the CF with $\sigma$; for a given value of $\alpha$, the CF decreases with increasing $\sigma$. The variation of CF with $\alpha$ is remarkable for the polytrope with indices n=0.5 and 1, slight for n=1.5, and vanishes for n=2.

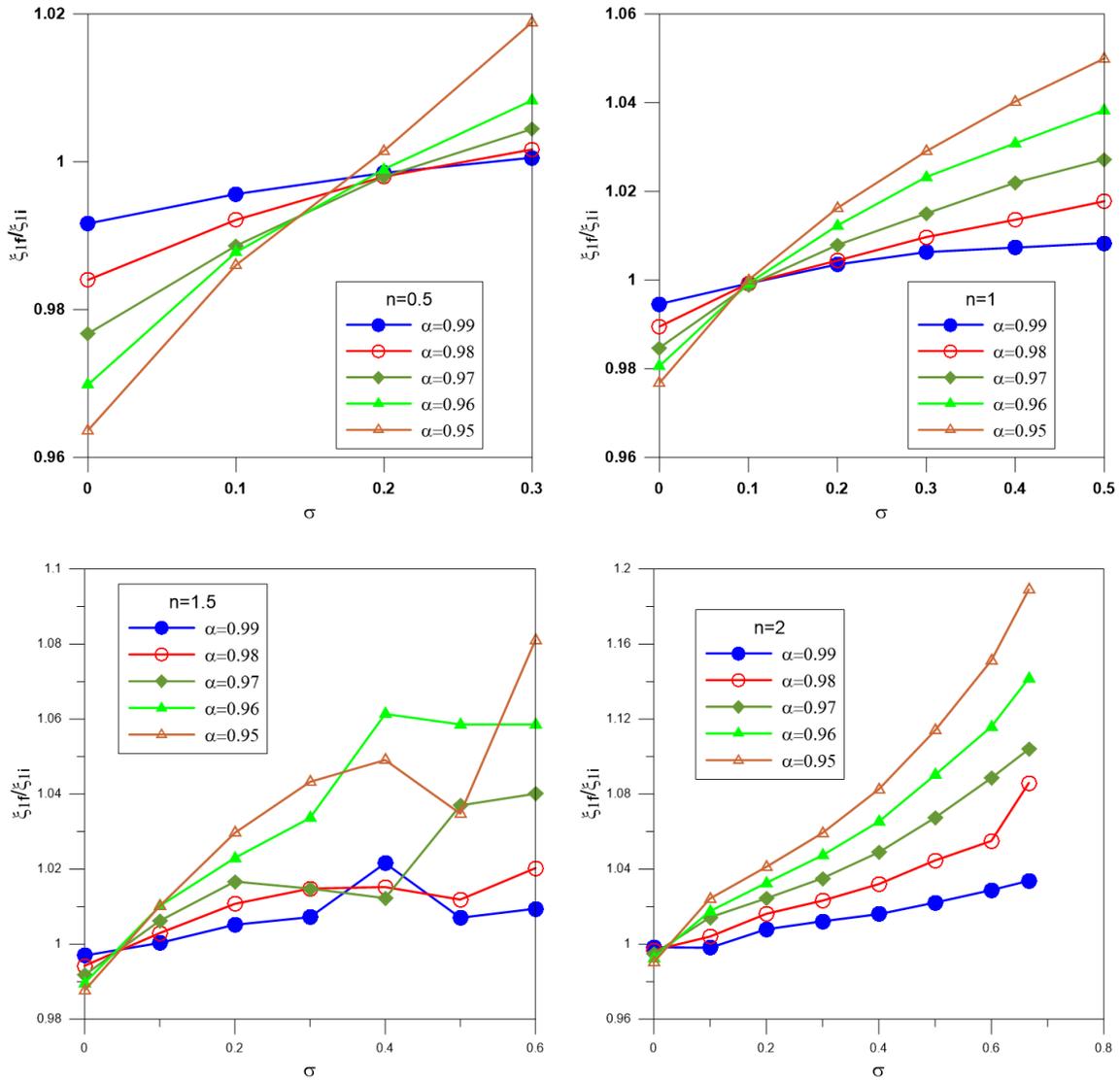

Figure 8: The ratio of the radius of the gas sphere computed from the fractional polytrope to that calculated from the integer polytrope; different colors denote different fractional parameters.

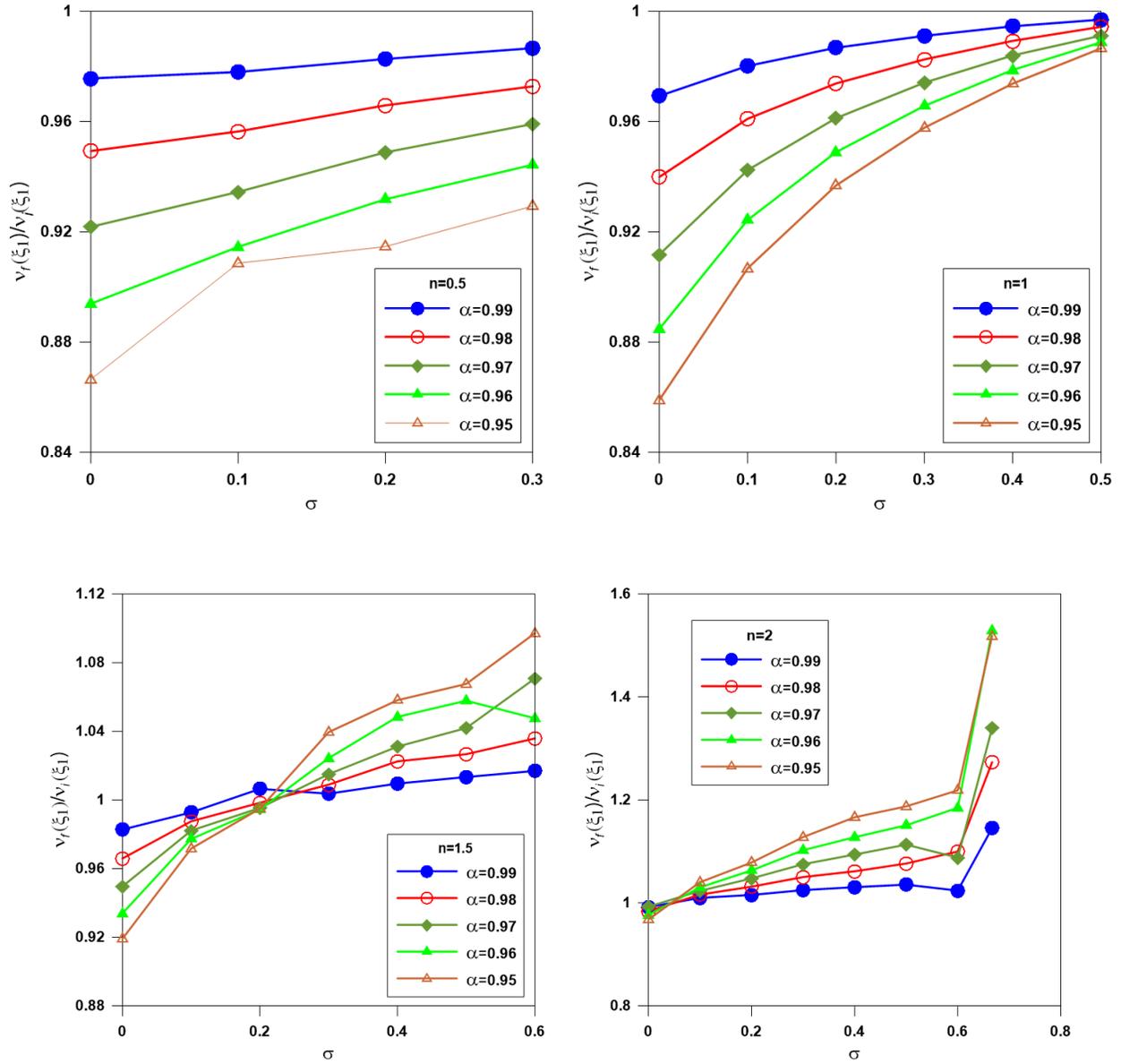

Figure 9: The ratio of the mass function computed from the fractional polytrope to that calculated from the integer polytrope; different colors denote different fractional parameters.

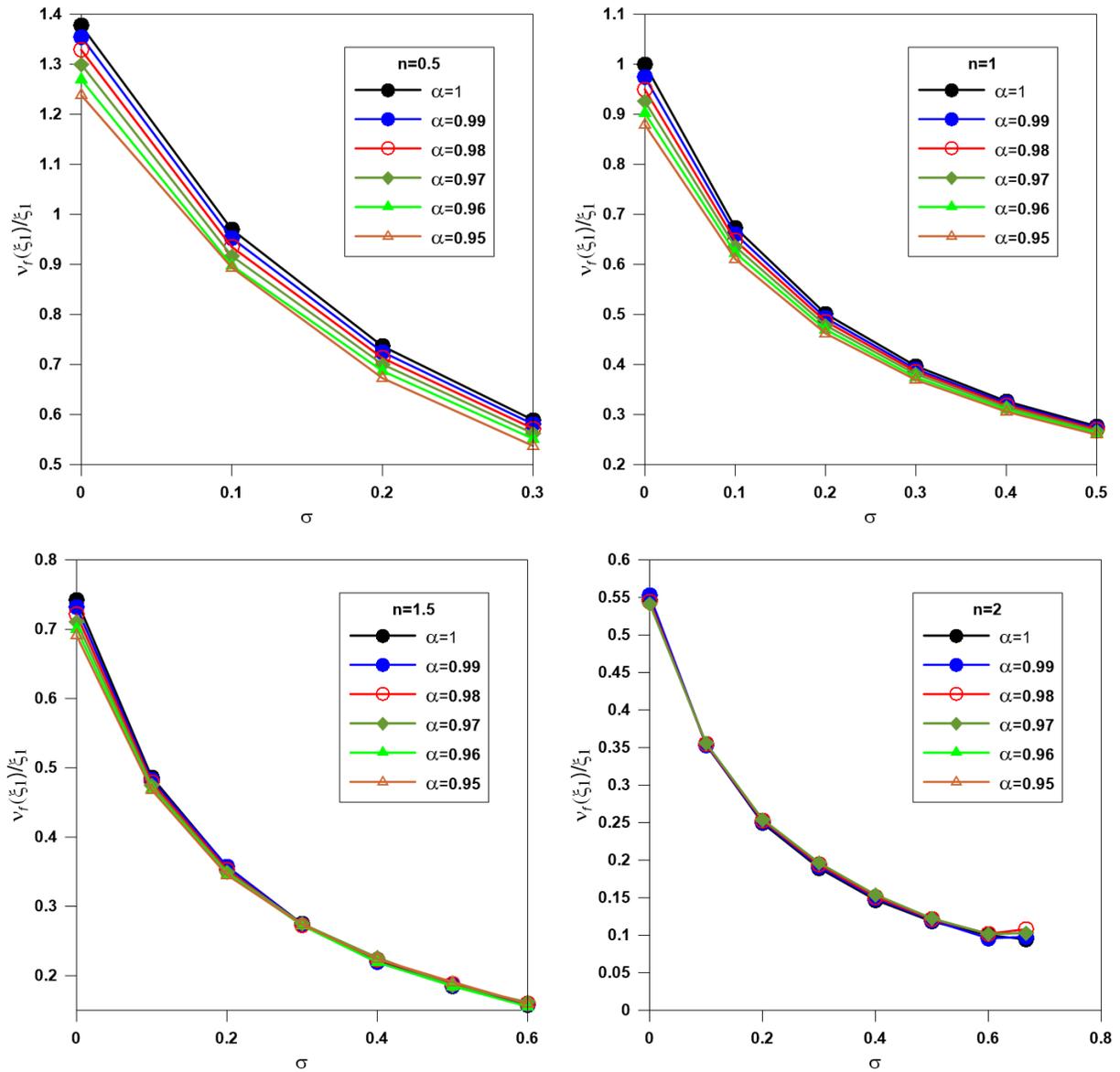

Figure 10: The ratio of the mass function to the zero of the Emden function ($\upsilon_f(\xi_1)/\xi_1$). Different colors denote different fractional parameters.

The polytrope with n=3 has a special interest when studying the relativistic effect in white dwarfs and neutron stars. Several authors find that general relativistic hydrostatic equilibrium yields a maximum mass slightly below the Chandrasekhar limiting mass [34-36]. If we consider relativistic electrons ($\rho_c >> 10^6$ g/cm$^3$, n=3), the radius and the mass of the polytrope could be calculated by Equations (80,81), where the constant $K$ is given by $K = 1.2435 \times 10^{15} / \mu_e^{4/3}$, $\mu_e = 2$ is the molecular weight as used in the Chandrasekhar calculations. Tables 13-18 list the mass limit ($M_{limit}$) and minimum radius ($R_{min}$) of the fractional gas sphere with polytropic index n=3. For the integer models with small relativistic parameters ($\alpha = 1$, $\sigma = 0.001$), we obtained a mass limit for the white dwarfs close to the Chandrasekhar limit. The situation is changed by reducing the fractional parameter; the mass limit grows and becomes $M_{limit}$=1.63348 $M_\odot$ at the $\alpha = 0.95$ and $\sigma = 0.001$.

Figure 11 plots the maximum mass ($M_{limit}$) and minimum radius ($R_{min}$) at several values of the relativistic and fractional parameters. Two features can be seen from the figure, the maximum mass, and the minimum radius decrease by reducing the fractional parameter and raising the relativistic parameter. The minimum radii decrease rapidly with increasing the relativistic parameter, whereas the difference when changing the fractional parameter is slight. For a given fractional parameter, the mass limit and the minimum radii computed at $\sigma = 0.3$ are reduced to about 30% and 5% of their values calculated at $\sigma = 0.01$ (i.e., the star lost about 70% of its mass and 95% of its radius). On the other hand, for a given relativistic parameter, the mass limit and minimum radii changes are about 20% and 8% for the mass limit and minimum radii (i.e., the star lost about 80% of its mass and 92% of its radius). This strong dependence of the mass limit and minimum radius on the relativistic and fractional parameters indicates the general relativity's importance when modelling the fractional polytropes.

The mass distribution of the observed white dwarfs significantly peaked at 0.6 M$_\odot$, with the bulk falling between 0.5 and 0.7 M$_\odot$ [38]. White dwarfs are known to have estimated masses as low as 0.17 M$_\odot$ [39] and as high as 1.33 M$_\odot$ [38]; the diameters are typically 0.8–2% of the Sun's radius [40]. The percentage of the ratio $R_{min}/R_\odot$ in the sixth column of Tables 13-18 supports these observational findings for fractional relativistic polytropes with $0 \leq \sigma \leq 0.01$.

The mass-radius and mass-density relations for white dwarfs are plotted in Figure 12. The mass goes rapidly high as the radius increase till $R_{min}$=2000 km, then the change with radius becomes very slow. Another feature from the figure is that most white dwarfs have radii smaller than 2000 km. The mass-density relation shows an interesting behaviour; the change in the mass limit is slow as the central density increases for $\log \rho_c = 9-13$ g/cm$^3$, and after this value, the change becomes faster.

Table 13: Mass limit and minimum radius of white dwarfs computed at $\alpha=1$.

| $\sigma$ | $\xi_1$ | $\upsilon(\xi_1)$ | $M_{\text{limit}}$ ($M_\odot$) | $R_{\min}$ (km) | $R_{\min}$ ($R_\odot$)% | $\log \rho_c$ |
|---|---|---|---|---|---|---|
| 0.001 | 6.864 | 2.01684 | 1.45516 | 12535.8 | 1.8011 | 9.7829 |
| 0.003 | 6.85 | 1.99952 | 1.44266 | 4161.58 | 0.5979 | 11.214 |
| 0.005 | 6.837 | 1.97217 | 1.42293 | 2487.48 | 0.3573 | 11.879 |
| 0.007 | 6.835 | 1.942892 | 1.4018 | 1775.73 | 0.2551 | 12.318 |
| 0.009 | 6.833 | 1.91368 | 1.38073 | 1380.32 | 0.1983 | 12.645 |
| 0.01 | 6.821 | 1.89885 | 1.37003 | 1237.93 | 0.1778 | 12.782 |
| 0.03 | 6.707 | 1.6245 | 1.17208 | 1196.89 | 0.1719 | 14.214 |
| 0.05 | 6.710 | 1.42668 | 1.02936 | 239.593 | 0.0344 | 14.879 |
| 0.07 | 6.723 | 1.26794 | 0.914824 | 171.801 | 0.0246 | 15.318 |
| 0.09 | 6.811 | 1.14184 | 0.823842 | 137.144 | 0.0197 | 15.645 |
| 0.1 | 6.872 | 1.07302 | 0.774188 | 125.651 | 0.0180 | 15.782 |
| 0.2 | 7.941 | 0.714168 | 0.515275 | 83.8917 | 0.0120 | 16.686 |
| 0.3 | 10.841 | 0.543307 | 0.391998 | 104.236 | 0.0149 | 17.214 |

Table 14: Mass limit and minimum radius of white dwarfs computed at $\alpha=0.99$.

| $\sigma$ | $\xi_1$ | $\upsilon(\xi_1)$ | $M_{\text{limit}}$ ($M_\odot$) | $R_{\min}$ (km) | $R_{\min}$ ($R_\odot$)% | $\log \rho_c$ |
|---|---|---|---|---|---|---|
| 0.001 | 6.887 | 2.11124 | 1.52327 | 12620 | 1.8132 | 9.782 |
| 0.003 | 6.868 | 2.08854 | 1.50689 | 4183.48 | 0.6010 | 11.214 |
| 0.005 | 6.857 | 2.07092 | 1.49418 | 2502.05 | 0.3594 | 11.879 |
| 0.007 | 6.564 | 2.02804 | 1.46324 | 1637.71 | 0.2353 | 12.318 |
| 0.009 | 6.855 | 2.06725 | 1.49153 | 1389.22 | 0.1996 | 12.645 |
| 0.01 | 6.831 | 2.07538 | 1.49739 | 1241.56 | 0.1783 | 12.782 |
| 0.03 | 6.772 | 1.7009 | 1.22721 | 610.102 | 0.0876 | 14.214 |
| 0.05 | 6.650 | 1.49064 | 1.0755 | 235.327 | 0.0338 | 14.879 |
| 0.07 | 6.938 | 1.33857 | 0.965783 | 182.965 | 0.0262 | 15.318 |
| 0.09 | 6.963 | 1.20855 | 0.871973 | 143.334 | 0.0205 | 15.645 |
| 0.1 | 7.097 | 1.1182 | 0.806786 | 134.013 | 0.0192 | 15.782 |
| 0.2 | 8.039 | 0.760164 | 0.548461 | 85.9751 | 0.0123 | 16.686 |
| 0.3 | 10.342 | 0.596392 | 0.430299 | 94.8607 | 0.0136 | 17.214 |

Table 15: Mass limit and minimum radius of white dwarfs computed at $\alpha=0.98$.

| $\sigma$ | $\xi_1$ | $\upsilon(\xi_1)$ | $M_{\text{limit}}$ ($M_\odot$) | $R_{\text{min}}$ (km) | $R_{\text{min}}$ ($R_\odot$)% | $\log \rho_c$ |
|---|---|---|---|---|---|---|
| 0.001 | 6.863 | 2.153 | 1.5534 | 12532.2 | 1.8006 | 9.7829 |
| 0.003 | 6.893 | 2.13234 | 1.53849 | 4213.99 | 0.6054 | 11.2142 |
| 0.005 | 6.884 | 2.112 | 1.52382 | 2521.8 | 0.3623 | 11.8798 |
| 0.007 | 6.874 | 2.09632 | 1.5125 | 1796.05 | 0.2580 | 12.3182 |
| 0.009 | 6.877 | 2.08976 | 1.50777 | 1398.15 | 0.2008 | 12.6456 |
| 0.01 | 6.858 | 2.08744 | 1.5061 | 1251.39 | 0.1797 | 12.7829 |
| 0.03 | 6.813 | 1.75646 | 1.26729 | 411.675 | 0.0591 | 14.2142 |
| 0.05 | 6.560 | 1.55204 | 1.1198 | 229.0 | 0.0329 | 14.8798 |
| 0.07 | 7.001 | 1.398787 | 1.00923 | 186.303 | 0.0267 | 15.3182 |
| 0.09 | 7.065 | 1.26883 | 0.915466 | 147.564 | 0.0212 | 15.6456 |
| 0.1 | 7.107 | 1.16447 | 0.84017 | 134.391 | 0.0193 | 15.7829 |
| 0.2 | 7.905 | 0.807264 | 0.582444 | 83.1328 | 0.0119 | 16.6860 |
| 0.3 | 10.937 | 0.641067 | 0.462532 | 106.09 | 0.0152 | 17.2142 |

Table 16: Mass limit and minimum radius of white dwarfs computed at $\alpha=0.97$.

| $\sigma$ | $\xi_1$ | $\upsilon(\xi_1)$ | $M_{\text{limit}}$ ($M_\odot$) | $R_{\text{min}}$ (km) | $R_{\text{min}}$ ($R_\odot$)% | $\log \rho_c$ |
|---|---|---|---|---|---|---|
| 0.001 | 6.929 | 2.20077 | 1.58786 | 12774.4 | 1.8354 | 11.2142 |
| 0.003 | 6.917 | 2.17451 | 1.56892 | 4243.39 | 0.6096 | 11.2142 |
| 0.005 | 6.904 | 2.15108 | 1.55201 | 2536.47 | 0.3644 | 11.8798 |
| 0.007 | 6.891 | 2.13112 | 1.53761 | 1804.95 | 0.2593 | 12.3182 |
| 0.009 | 6.880 | 2.11583 | 1.52658 | 1399.37 | 0.2010 | 12.6456 |
| 0.01 | 6.878 | 2.11088 | 1.52301 | 1258.7 | 0.1808 | 12.7829 |
| 0.03 | 6.819 | 1.80284 | 1.30076 | 412.4 | 0.0592 | 14.2142 |
| 0.05 | 6.766 | 1.60963 | 1.16136 | 243.609 | 0.0350 | 14.8798 |
| 0.07 | 7.112 | 1.4582 | 1.0521 | 192.258 | 0.0276 | 15.3182 |
| 0.09 | 7.119 | 1.32488 | 0.955906 | 149.828 | 0.0215 | 15.6456 |
| 0.1 | 7.031 | 1.19663 | 0.863373 | 135.3 | 0.0194 | 15.7829 |
| 0.2 | 7.294 | 0.84647 | 0.610731 | 70.7783 | 0.0101 | 16.6860 |
| 0.3 | 10.549 | 0.707563 | 0.510509 | 98.6961 | 0.0141 | 11.2142 |

Table 17: Mass limit and minimum radius of white dwarfs computed at $\alpha=0.96$.

| $\sigma$ | $\xi_1$ | $\upsilon(\xi_1)$ | $M_{\text{limit}}$ ($M_\odot$) | $R_{\text{min}}$ (km) | $R_{\text{min}}$ ($R_\odot$)% | $\log \rho_c$ |
|---|---|---|---|---|---|---|
| 0.001 | 6.943 | 2.23836 | 1.61499 | 12826 | 1.8428 | 9.782 |
| 0.003 | 6.952 | 2.21357 | 1.5971 | 4286.44 | 0.6158 | 11.214 |
| 0.005 | 6.934 | 2.18773 | 1.57846 | 2558.56 | 0.3676 | 11.879 |
| 0.007 | 6.92 | 2.16509 | 1.56212 | 1820.17 | 0.2615 | 12.318 |
| 0.009 | 6.929 | 2.14891 | 1.55045 | 1419.37 | 0.2039 | 12.645 |
| 0.01 | 6.898 | 2.1370 | 1.54185 | 1266.03 | 0.1819 | 12.782 |
| 0.03 | 6.813 | 1.8439 | 1.33038 | 411.675 | 0.0591 | 14.214 |
| 0.05 | 6.784 | 1.65237 | 1.19219 | 244.906 | 0.0351 | 14.879 |
| 0.07 | 7.162 | 1.51095 | 1.09016 | 194.971 | 0.0280 | 15.318 |
| 0.09 | 7.237 | 1.38185 | 0.99701 | 154.836 | 0.0222 | 15.645 |
| 0.1 | 7.119 | 1.23351 | 0.889982 | 134.846 | 0.0193 | 15.782 |
| 0.2 | 7.063 | 0.886431 | 0.639563 | 66.3662 | 0.0095 | 16.686 |
| 0.3 | 10.801 | 0.720947 | 0.520166 | 103.468 | 0.0148 | 17.214 |

Table 18: Mass limit and minimum radius of white dwarfs computed at $\alpha=0.95$.

| $\sigma$ | $\xi_1$ | $\upsilon(\xi_1)$ | $M_{\text{limit}}$ ($M_\odot$) | $R_{\text{min}}$ (km) | $R_{\text{min}}$ ($R_\odot$)% | $\log \rho_c$ |
|---|---|---|---|---|---|---|
| 0.001 | 6.909 | 2.26399 | 1.63348 | 12700.7 | 1.8248 | 9.7829 |
| 0.003 | 6.976 | 2.24677 | 1.62105 | 4316.09 | 0.6201 | 11.2142 |
| 0.005 | 6.973 | 2.22179 | 1.60303 | 2587.43 | 0.3717 | 11.8798 |
| 0.007 | 6.95 | 2.19614 | 1.58452 | 1835.99 | 0.2637 | 12.3182 |
| 0.009 | 6.954 | 2.1769 | 1.57064 | 1429.64 | 0.2054 | 12.6456 |
| 0.01 | 6.931 | 2.16455 | 1.56173 | 1278.17 | 0.1836 | 12.7829 |
| 0.03 | 6.864 | 1.88557 | 1.36045 | 417.861 | 0.0600 | 13.7844 |
| 0.05 | 6.851 | 1.6961 | 1.22374 | 249.768 | 0.0358 | 14.879 |
| 0.07 | 7.195 | 1.55853 | 1.12449 | 196.771 | 0.0282 | 15.3182 |
| 0.09 | 7.295 | 1.43221 | 1.03335 | 157.328 | 0.0226 | 15.6456 |
| 0.1 | 7.452 | 1.27312 | 0.918561 | 147.756 | 0.0212 | 15.7829 |
| 0.2 | 7.876 | 0.961846 | 0.693976 | 82.5239 | 0.0118 | 16.6860 |
| 0.3 | 10.633 | 0.804666 | 0.58057 | 100.274 | 0.0144 | 17.2142 |

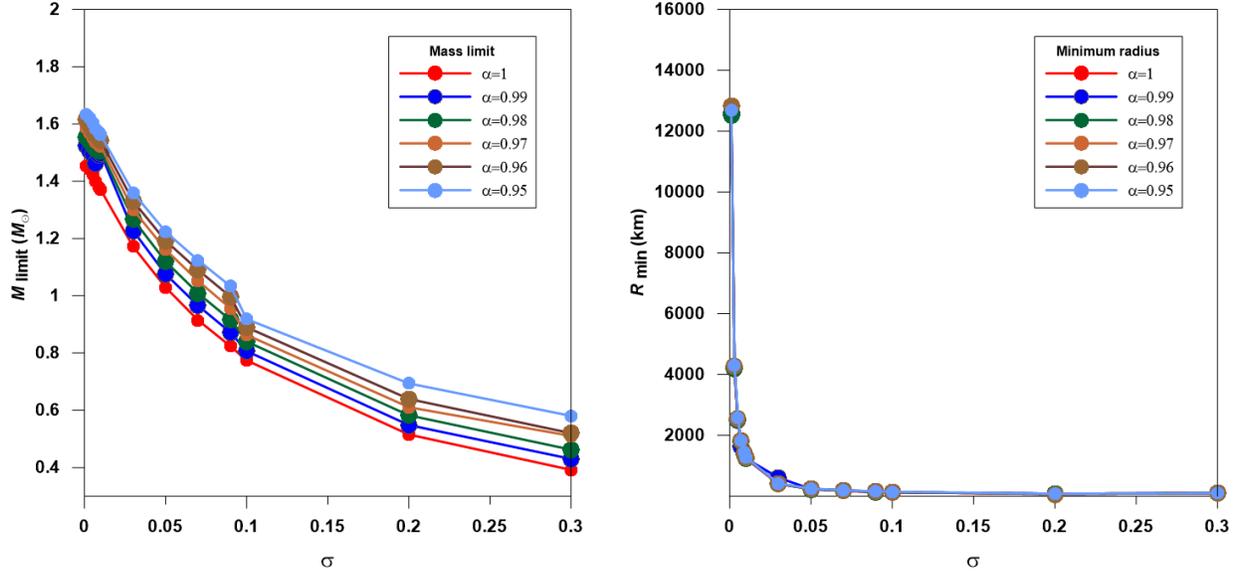

Figure 11: The distribution of the mass limit (left panel) and the minimum radius (right panel) with the relativistic parameter for fractional white dwarf stars modeled with n=3 polytrope.

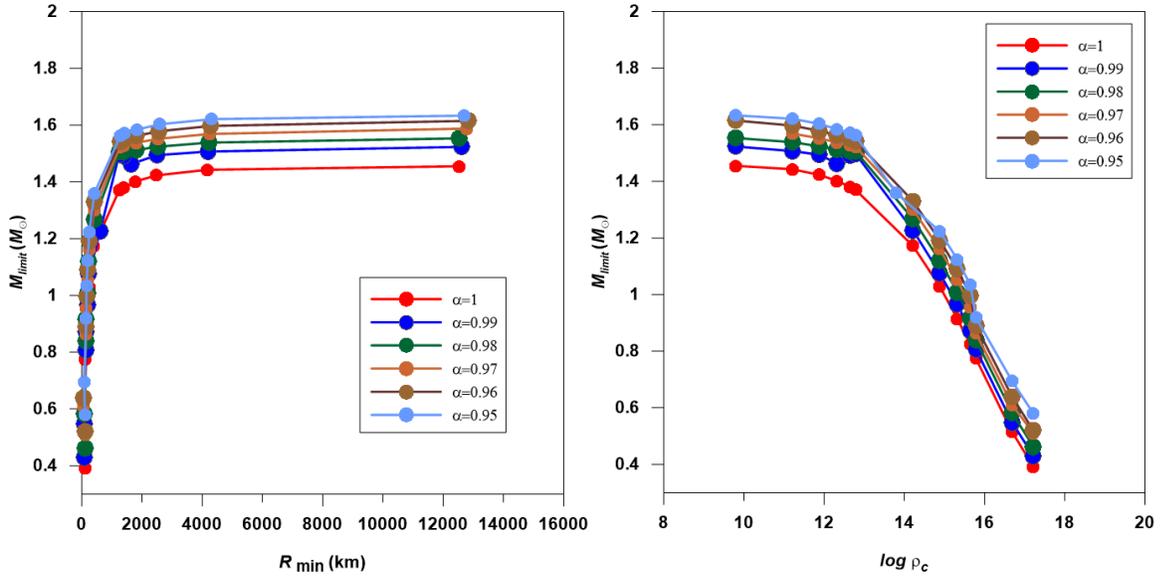

Figure 12: The mass-radius and mass-density relations for the fractional white dwarfs modeled with n=3 polytrope.

## 6. Conclusions

In the present paper, we introduce an approximate solution to the fractional Lane-Emden equation via accelerated series expansion by constructing a recurrence relation for the Emden and mass functions using the fractal index. The findings demonstrate that the accelerated series solution converges everywhere for the polytropic index for a range of the polytropic indexes n=0-3. We found that the Emden function's zeroth and the mass function vary with fractional and relativistic parameters; the volume of the gas sphere changes with changing the fractional and relativistic parameters. We calculated the compactification factor by dividing the mass function by the zero of the Emden function. With rising relativistic parameters, the compactification factor falls. The variation of the fractional parameter of the compactification factor is significant for the polytrope with indices n=0.5 and 1, modest for n=1.5, and disappears for n=2. As the stellar matter density is more concentrated in the star's centre as the relativistic parameter rises (the sound velocity is not small relative to the speed of light) [10], the relativistic fermi gas (n = 3) makes this effect considerably more pronounced. At the star's core, where the metric functions depend on the mass to radius ratio, which gauges the star's compactness, it will significantly distort spacetime.

We calculated the maximum mass and the corresponding minimum radius of white dwarfs for a wide range of relativistic parameters (for the ultra-relativistic polytrope n=3). The results revealed that the maximum mass decreases as the fractional parameter decreases, and we found that the maximum mass limit of fractional white dwarfs is changed by changing both the relativistic and fractional parameters. For a particular fractional parameter, the mass limit and the minimum radii estimated at $\sigma = 0.3$ are lowered to approximately 30% and 5% of their values calculated at $\sigma = 0.01$. Additionally, the change mass limit and minimum radii calculated at $\alpha = 0.99$ are roughly 20% and 8% smaller than those calculated at $\alpha = 0.95$ for a particular relativistic parameter. The significance of integrating relativistic corrections in the calculations of the polytropic gas sphere and the mass limit of the white dwarfs is highlighted by this striking difference. Besides, the fractional and relativistic parameters on the mass and radius of the polytrope changed the physical conditions inside the sphere; we found that the present fractional relativistic polytropes will be helpful for detailed modeling of the structure of white dwarfs and neutron stars, which will be the subject of a future paper.

**Competing interests**

The author(s) declare no competing interests.

**Acknowledgment:** This paper is based upon work supported by Science, Technology & Innovation Funding Authority (STDF) under the grant number 37122. The authors thank the editor and reviewers for their valuable comments.

Contributions

M.A. and M.N. conceived the idea of the study. E.A. and T.K. conducted the analytical solution. M.N. performed the analysis of the results. M.A. performed the coding and analytical calculations. The first draft of the manuscript was written by M.N, M.A., E.A., M.B., and K.G. T.K. K.G. contributed to structuring and editing the manuscript and approved the submitted version.

Data availability

The datasets used and/or analysed during the current study are available from the corresponding author upon reasonable request.

[7] Linares L., Malherio S. and S. Ray S. , The Importance of the Relativistic Corrections in Hyperon Stars, Int. J. Mod. Phys. D, 13, 1355, (2004).

[8] Abdul Aziz, Saibal Ray, Farook Rahaman, B.K. Guha, Neutron star under homotopy perturbation method, Annals of Physics, Volume 409, 2019, 167918.

[9] Tooper R., General Relativistic Polytropic Fluid Spheres, Astrophys. J., 140, 434, (1964).

[10] Nouh M.I., Saad A.S., "A New Analytical Solution to the Relativistic Polytropic Fluid Spheres ", International Review of Physics, Int. Rev. Phys.7, 1, (2013).

[11] Hunter C., Series solutions for polytropes and the isothermal sphere, Monthly Notices of the Royal Astronomical Society, Volume 328, Issue 3. Sci.328, 839-847, (2001).

[12] Nouh, M. I., Accelerated power series solution of the polytropic and isothermal gas spheres, New Astron. 9, 467, (2004).

[13] Debnath, U. ; Jamil, M. and Chattopadhyay, S., Fractional Action Cosmology: Emergent, Logamediate, Intermediate, Power Law Scenarios of the Universe and Generalized Second Law of Thermodynamics, International Journal of Theoretical Physics, Volume 51, Issue 3, pp.812-837, (2012).

[14] Jamil, M., Momeni, D., Rashid, M. A.: Fractional action cosmology with power law weight function. J. Phys. Conf. Ser 2012, 354: 01, (2008).

[15] El-Nabulsi, R. A., The fractional white dwarf hydrodynamical nonlinear differential equation and emergence of quark stars, Applied Mathematics, and Computation, 218, 2837, (2011).

[16] Yousif, E., Adam, A. Hassaballa, A. and Nouh, M. I., Conformable Isothermal Gas Spheres, New Astronomy, 101511, (2021).

[17] Abdel-Salam E. AB., Nouh M.I. Approximate Solution to the Fractional Second-Type Lane-Emden Equation. Astrophysics 59, 398–410, (2016).

[18] Nouh, M.I., Abdel-Salam, E. A-B., Analytical solution to the fractional polytropic gas spheres, European Physics Journal Plus, 133, 149, (2018).

[19] Abdel-Salam, E.A-B, Nouh, M.I., Conformable Polytropic Gas Spheres, New Astronomy 2020; 76, 101322, (2020).

[20] Jumarie G., Laplaces Transform of Fractional Order via the Mittag Leffler Function and Modified Riemann-Liouville Derivative, Applied Mathematics Letters, Vol. 22, No. 11 pp. 1659-1664, (2009).